\documentclass[superscriptaddress,secnumarabic,amssymb,amsmath,nobibnotesquantum universes,aps,prd,nofootinbib ]{revtex4}
\usepackage{color}
\usepackage{graphicx}
\usepackage{dcolumn}
\usepackage{bm}

\newcommand{\beq}{\begin{equation}}
\newcommand{\eeq}{\end{equation}}
\newcommand{\bey}{\begin{eqnarray}}
\newcommand{\eey}{\end{eqnarray}}

\begin{document}

\title{Wormholes in $\kappa(R,T)$ gravity}

\author{Susmita Sarkar}
 \email{ susmita.mathju@gmail.com}
\affiliation {Department of Mathematics, Jadavpur University, Kolkata-700032, India}

\author{Nayan Sarkar}
 \email{ nayan.mathju@gmail.com}
\affiliation {Department of Mathematics, Jadavpur University, Kolkata-700032, India}

\author{Farook Rahaman}
\email{rahamank@associates.iucaa.in}
\affiliation{
Department of Mathematics, Jadavpur University, Kolkata 700 032, West Bengal,
India.}

\author{Y  Aditya}
\email{yaditya2@gmail.com}
\affiliation{
Department of Mathematics, ANITS(A), Visakhapatnam 531 162, A.P., India
.}

\date{\today}
\begin{abstract}
The wormhole solution could be found by solving the Einstein field equations with violating the  null energy condition (NEC). We represent  wormhole solutions in $\kappa(R,T)$ gravity in two different ways.  At first, we find the shape
 function by considering a redshift function and linear equation of state (EoS). The solution represents a wormhole  for the real feasible matter. In the second part, we consider four pairs of  two redshift functions and two shape functions and analyze the obtained solutions. Some of  the models  suggest that for particular values of the parameters,  the   existence of wormholes are supported by an arbitrarily small quantity of exotic matter.
\end{abstract}

\maketitle


\section{\label{sec1} Introduction}

In general relativity, one of the noteworthy features is the possible existence of hypothetical geometry consisting nontrivial topological structure. This topological feature of space-time as the solution of Einstein's general relativity field equations called as wormholes.
Wormhole (WH) is a hypothetical path which connects two different space-times. In recent times it becomes a very interesting research object in theoretical Astrophysics as if traversable WH exits then time machine can be constructed \cite{mv95}. Flamm \cite{fl16} has first introduced the concept of WH and discussed 2D embedding diagram of Schwarzschild WH. Later a similar attempt has been made by Einstein and Rosen \cite{en35} and proposed the non-traversable Schwarzschild WH solutions. Also, this doesn't allow a two-way transmission between two different regions of the space-time that leads to the contraction of WH throat.

The concept of Lorentzian traversable WH proposed by Morris and Thorne \cite{ms88} has attracted many researchers. In this case, the throat of WH is threaded by a variety of exotic matter which causes repulsion
against the collapse of WH throat. Due to this condition of the traversable WH, it must satisfy the flare-out condition to preserve the geometry. Also, because of the absence of the event horizon in this traversable WH, the observers can traverse freely from one place of the universe to another place. Morris and Thorne  \cite{ms88} have also found that a kind of exotic matter is responsible for traversable WH which violates the null energy condition(NEC) and weak energy condition (WEC). NEC is the weakest one whose violation gives rise to the violation of other energy conditions. Thus, it is a fascinating challenge to search for a realistic model that minimize the usage of exotic matter or satisfies the energy conditions.

One of the most striking research areas over the last two decades has been the scenario of the accelerated expansion of our universe. It is thought that the reason behind this cosmic accelerated expansion of the universe is some exotic fluid or unidentified energy with huge negative pressure dubbed as dark energy(DE). Observational data from various experiments such as high redshift supernovae, galaxy clustering and cosmic microwave background anisotropy support this accelerated expansion of our universe \cite{ag98,sp99,ds03,mt04}. According to the latest observations, our universe constitutes about 71.4\% of DE, 24\% dark matter and 4.6\% ordinary matter. After many efforts by astronomers to find the nature and composition of DE but till we have just known that it has some repulsive force or negative pressure. To describe the DE, one can use the equation of state(EoS) parameter $\omega=p/\rho$, where $p$ and $\rho$ denote the pressure and energy density of the DE. Vacuum energy ($\omega=-1$) is the most convenient explanation for the DE that can be able to push matter apart \cite{pj03}, but it faces some theoretical problems like the fine tuning and cosmic coincidence.

Modified theories of gravitation are the another approaches to address the DE problem. These modified theories can be obtained by generalizing or modifying the Einstein-Hilbert action in standard GR. This idea gives rise to various alternative theories of gravitation like $f(\tau)$ gravity \cite{rf07,gr09,ev10}, Gauss-Bonnet gravity \cite{sm05, gc06}, $f(R)$ gravity \cite{sc03, sn03} and $f(R,T)$ gravity \cite{th11} etc (where $R$ is the curvature scalar, $\tau$ denotes the torsion scalar, $T$ is the trace of the energy momentum tensor). For an exhaustive study of modified gravities, the reader may see the reviews \cite{sc11,sn11,sn17}. On the other hand, the models in these modified theories may overcome the violation of energy conditions and lead to the WHs threaded by ordinary matter. Basically, in modified theories, the effective stress-energy tensor (extra terms or modified terms) is responsible for the violation of energy conditions while ordinary matter satisfies these conditions. Harko et al. \cite{th13} have presented theoretical WH geometries that can be obtained without the presence of exotic matter but are validated in the context of $f(R)$ modified gravity with higher order curvature terms. Rahaman et al. \cite{fr14} have derived some new exact solutions of static noncommutative WHs in $f(R)$ theory of gravity supported by the Lorentzian density distribution. Lobo and Oliveira \cite{fs09} have constructed traversable WH geometries within the framework of $f(R)$ theories of gravity. Mehdizadeh et al. \cite{mr15} have explored traversable WHs in Einstein-Gauss-Bonnet which satisfies the WEC. Zubair et al. \cite{mz16} have discussed static spherically symmetric WHs in $f(R,T)$ modified theory of gravity with matter contents as isotropic, anisotropic and barotropic fluids, and shown that the WH solutions can be obtained without exotic matter in few regions of space-time. Moraes et al. \cite{ph17} have studied general solutions for WHs with constant throat radius in $f(R,T)$ gravity. Sharif and Ikram \cite{ms15} have studied the traversable WH solutions in the framework of $f(G)$ gravity (where $G$ is the Gauss-Bonnet curvature invariant) and found that the effective stress-energy tensor violates the NEC throughout the WH throat. {Sarkar et al. \cite{sa18} have discussed WH in a modified theory of gravity in the background five dimensional Kaluza–Klein cosmology. Samanta et al.   \cite{sam18} have presented a comparative study of traversable WHs with exponential shape function in $f(R)$ and $f(R,T)$ modified gravities and in general relativity. Rosa et al. \cite{ro18} have explored WH solutions in a generalized hybrid metric-Palatini matter theory. Elizalde and Khurshudyan \cite{el18} have investigated WH formation in $f(R,T)$ gravity with radial pressure which admits an equation of state parameter of varying Chaplygin gas. Zubair et al. \cite{zu18} have studied static spherically symmetric WHs in generalized $f(R,\phi)$ gravity.  Jusuf et al. \cite{ju18} have discussed the deflection of light by black holes and WHs in the context of massive theory of gravity. Godani and Samanta \cite{go19} have obtained traversable WHs with two different shape functions in $f(R)$ gravity. }

The field equations obtained in the above modified theories are much more complicated than those of GR and in the process of modification some of the beauty of the original theory is also lost. In view of this, very recently, Gines R. Perez Teruel \cite{gr18} has introduced a new modified theory named as $\kappa(R,T)$ gravity. In the formulation of this theory he doesn't begin with the standard modified theory of gravity approach. Rather, he added the possible source terms directly to GR field equations by following Maxwell's and Einstein's original approaches. In particular, $\kappa(R,T)$ gravity is based on a natural extension of GR, where the modified field equations are obtained by adding the terms that only include the scalar curvature $R$ and the trace $T$ of the stress-energy tensor. So far, no one has investigated WH solutions in the context of $\kappa(R,T)$ theory of gravity, hence it would be interesting to study WH solutions in the framework of $\kappa(R,T)$ modified theory of gravity.

{The innovation points of this work are: (i) For the choice of the redshift function and linear  equation of state(EoS), the obtained solution represents the wormhole structure with a peculiar property that the original matter distribution is of real feasible matter and it provides the fuel to sustain the wormhole in $\kappa(R,T)$  gravity. (ii) The solutions corresponding to four pairs of two  shape functions and two redshift functions also represent the wormholes and here some of  the models  suggest that for particular values of the parameters,  the   existence of wormholes are supported by an arbitrarily small quantity of exotic matter.}

The context has been designed as follows: In Sec. II we have described $\kappa(R,T)$ gravity and written down the Einstein field equations in this gravity also mentioned the required conditions on the redshift and shape function to present the wormhole structure. The shape function and the corresponding solution are analyzed by redshift function and EoS taking into the account in Sec.III. In Sec.IV we have shown the wormhole solutions by considering different redshift and shape functions. Finally, the discussion has been made in Sec.V.

\begin{figure}[h]
\centering 
\includegraphics[width=6.5cm]{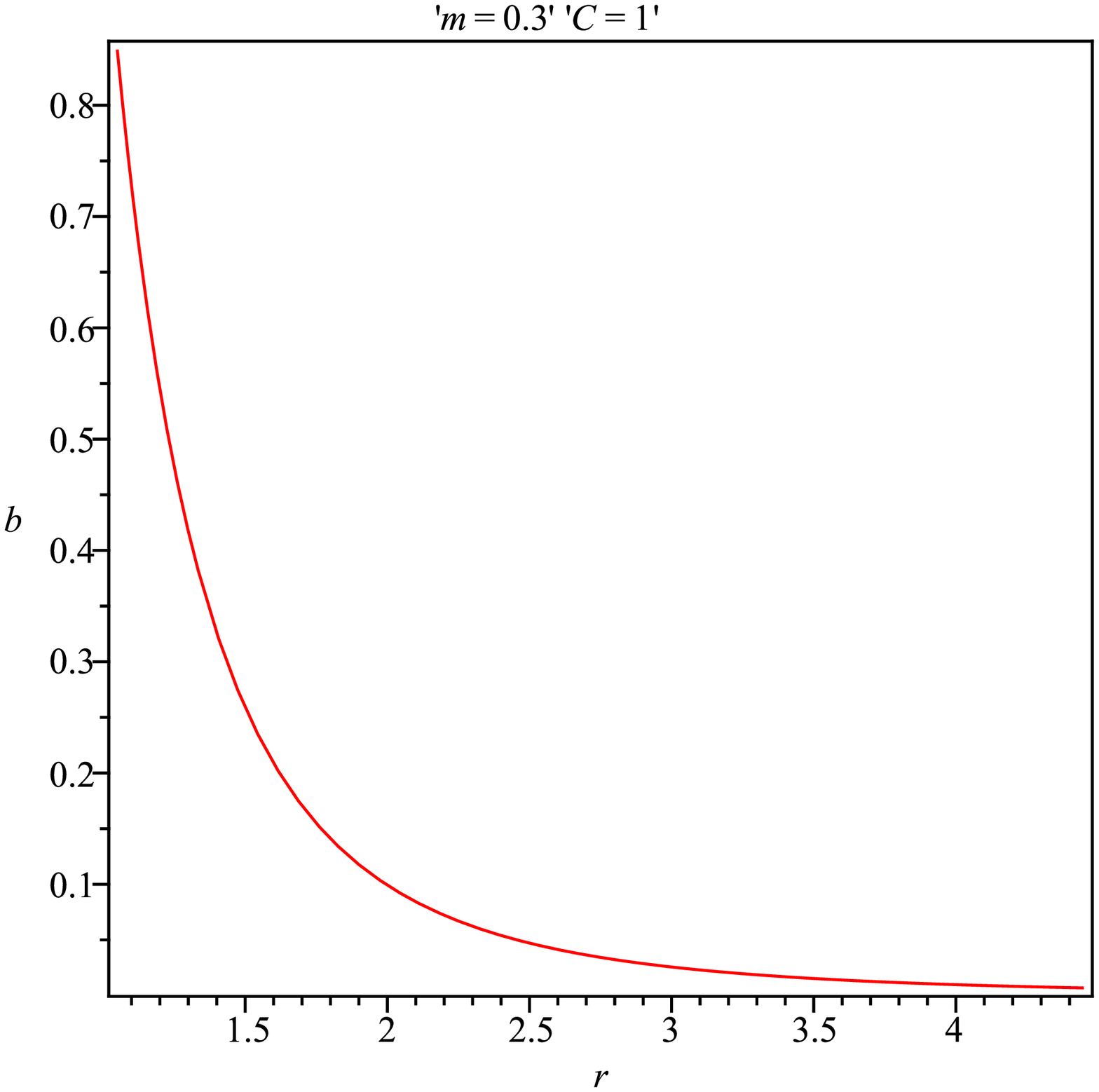}
\includegraphics[width=6.5cm]{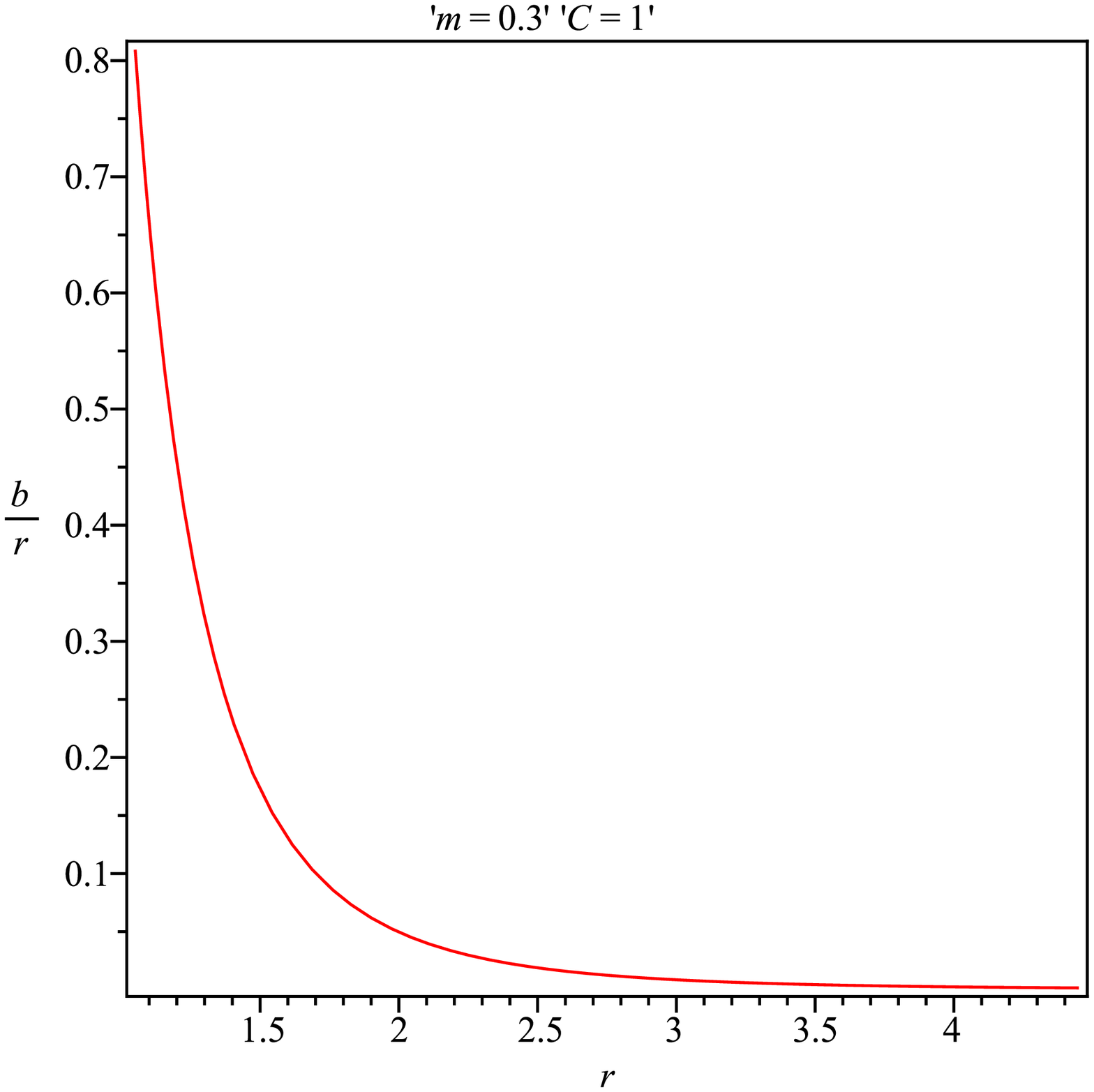}
\caption{ The shape function(Left) and  $\frac{b(r)}{r}$(Right) are plotted  against  the radial coordinate $r ( >$ the throat radius)  corresponding to $m = 0.3, C = 1$.}\label{fig1}
\end{figure}

\section{Einstein's field equations in $\kappa(R,T)$ gravity }
 Here, we will present a brief introduction to $\kappa(R,T)$ gravity, the corresponding field equations and required conditions on redshift and shape functions.

{  Various modified theories of gravitations are proposed in view of the important degree of arbitrariness inherent in the choice of the gravity Lagrangian. Many of these theories are so similar that it is difficult to differentiate one from the other. The Lagrangian formalism has undoubted advantages at the level of symmetries implementation and conservation-laws derivation, but possible theoretical alternatives to standard Lagrangian theories also deserve consideration. In this point of view, the importance of Non-Lagrangian theories in other branches of theoretical physics such as quantum field theory is being acknowledged in the last years. Also, these theories offer new opportunities in the search of new types of invariants.
	
Very recently, Teruel \cite{gr18} has formulated an example of a Non-Lagrangian modified theory of gravitation, namely $\kappa (R,T)$ gravity, inspired by Maxwell's approach to Electrodynamics, adding new possible source terms directly in the field equations}. The field equations in $\kappa(R,T)$ modified gravity are obtained by adding new possible source terms directly to GR field equations as\cite{gr18}
\begin{equation}
R_{ij}-\frac{1}{2}R g_{ij}-\Lambda g_{ij}=\kappa(R,T)T_{ij}\label{eq}
\end{equation}

where $g_{ij}$ is the metric potential, $R_{ij}$ is the Ricci tensor, $\Lambda$ is a cosmological constant, $T_{ij}$ is the energy-momentum tensor of the matter source, and $\kappa(R,T)$ corresponds to the Einstein gravitational constant and it is proposed as a function of the traces $T = g_{ij}T^{ij}$, and Ricci scalar $R = g_{ij}R^{ij}$. Clearly, the gravitational constant $\kappa$ depends on the scalars, so we can explore the possibility of a varying gravitational constant, i.e. generalization of the original Einstein's gravitational constant (not at the level of an action functional). A varying gravitational constant in the action leads to a Brans-Dicke scalar-tensor theory type \cite{cb61, cb05} with entirely different field equations from Eq.(\ref{eq}). Since the left hand side of the field  Eq.(\ref{eq}) is divergence free, we have

\begin{eqnarray}
\label{m1e}\nabla^j\left(\kappa(R,T) T_{ij}\right)&=& 0
\end{eqnarray}
Then, these field equations imply the non-covariant conservation of $T_{ij}$ that can be expressed as
\begin{eqnarray}
\label{m1f}\nabla^jT_{ij}&=& -\frac{\nabla^j\kappa(R,T)}{\kappa(R,T)}T_{ij}.
\end{eqnarray}
Teruel\cite{gr18} has proposed and analyzed some cosmological implications of two particular models. The first one is proposed by setting, $\kappa(T) = 8\pi G -\lambda T$, and corresponds to a matter-matter coupling. The second model is characterized by a gravitational constant that varies as $\kappa^{'}(R) = 8\pi G +\alpha R$, which will provide a coupling between matter and curvature terms.

Let us consider a Morris-Thorne wormhole, which is represented by a static spherically symmetric  metric in $Schwarzschild$ co-ordinate ($t$, $r$, $\theta$, $\phi$) as
\begin{equation}
ds^2 = e^{2f(r)}dt^2-\left(1-\frac{b(r)}{r}\right)^{-1}dr^2-r^2(d\theta^2 + sin^2\theta d\phi^2)\label{metric}
\end{equation}

\begin{figure}[h]
\centering
\includegraphics[width=6.5cm]{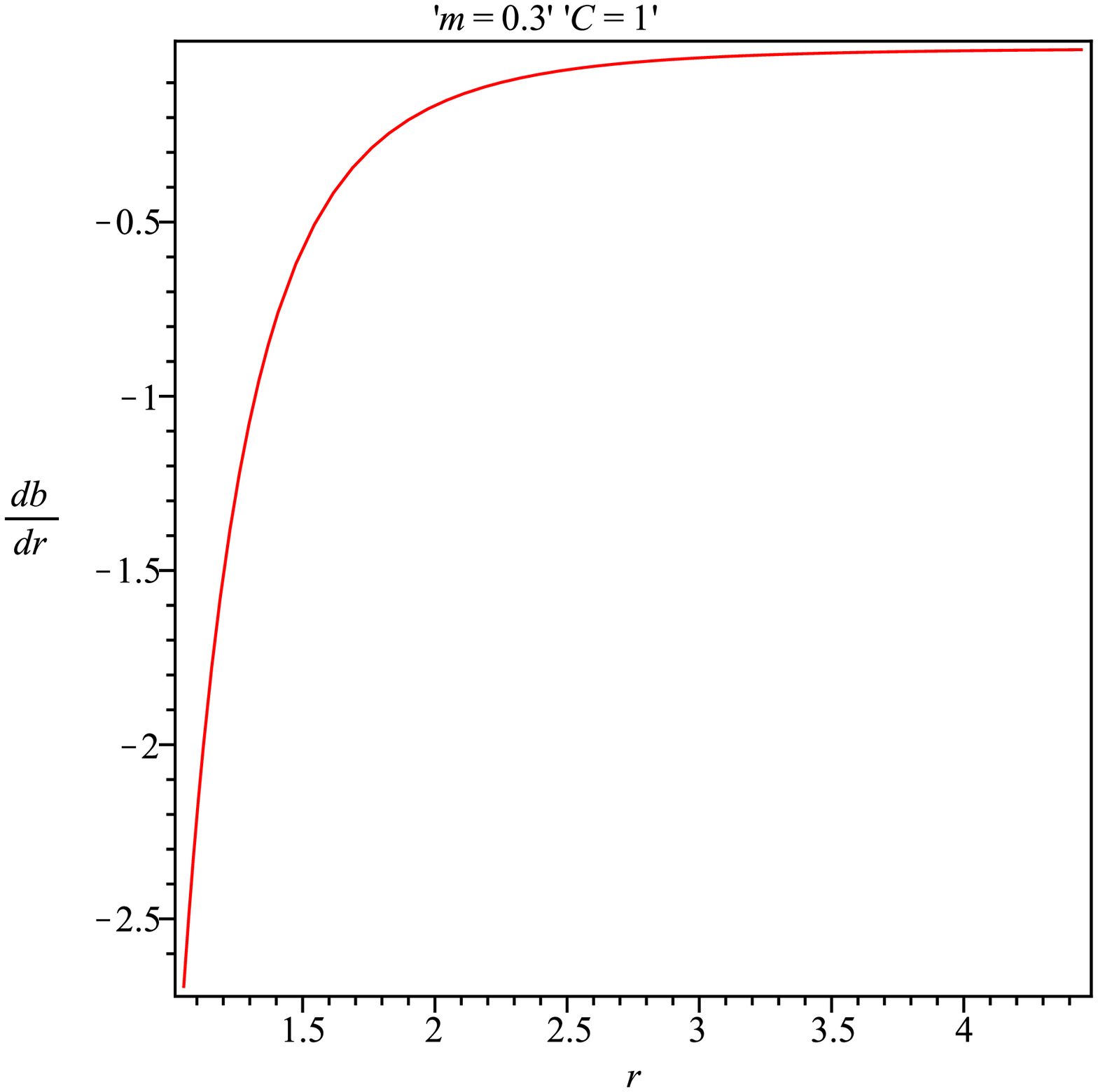}
\includegraphics[width=6.5cm]{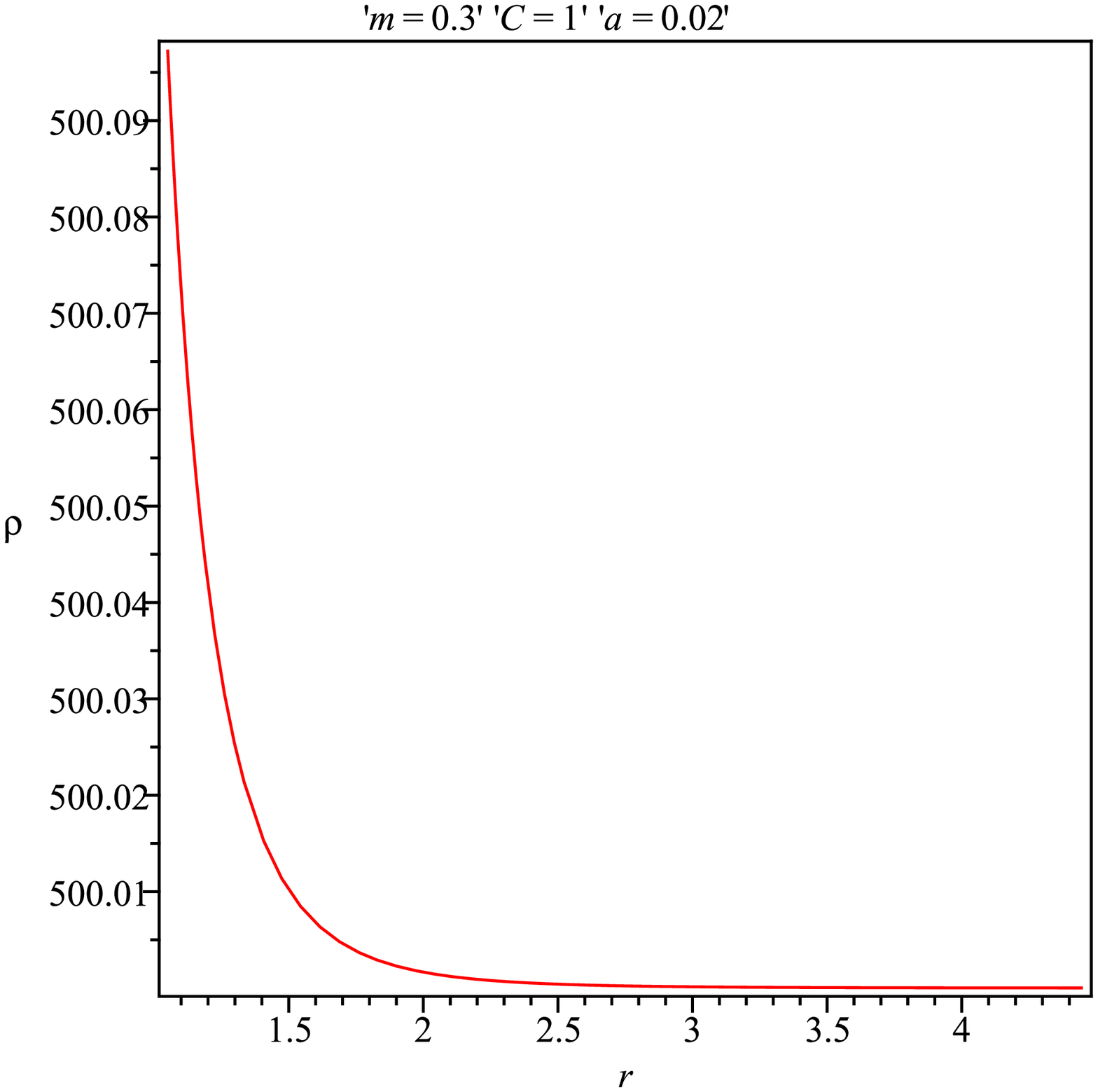}
\caption{ $\frac{db(r)}{dr}$(Left) and the original matter density(Right) are   are plotted  against  the radial coordinate $r ( >$ the throat radius) corresponding to $m = 0.3, C = 1$ and $m = 0.3, C = 1, a = 0.02$, respectively.}\label{fig2}
\end{figure}

where  $f(r)$ and $b(r)$ are  the functions of the radial coordinate $r$, called redshift function and shape function, respectively. The minimum radius $r= r_0$ in the metric coefficient $g_{rr}$ is termed as the throat radius of the wormhole, where $b(r_0) = r_0$ . Also, the absence of the event horizon is required for a traversable wormhole. \\
 Therefore, to represent a wormhole structure the redshift and shape functions must satisfy the following conditions \cite{fr06} :
 \begin{itemize}
 	\item[(i)] For avoiding an event horizon the redshift function  $f(r)$ should be finite.
 	\item[(ii)] The shape function  $b(r)$ should  satisfy the flare-out condition at the throat $r = r_0$ i.e.  $b^{\prime}(r_0) < 1$.
 	\item[(iii)] $b(r)$ should be less than r for $r>r_0$.
 	\item[(iv)] The shape function should be asymptotically flat i.e. $\frac{b(r)}{r}\rightarrow$ 0 as r$\rightarrow \infty$
 \end{itemize}

  To know the nature of matter that may support the wormhole we need to solve the Einstein field equations. The Einstein field equations in $\kappa(R,T)$ gravity  for the line element(\ref{metric}) along with $\Lambda = 0$ and $\kappa(R,T) = 8\pi-\lambda T$ $(G = c = 1)$  are:

\begin{eqnarray}
8\pi \rho(r)\left[1-a(\rho(r)-3p(r))\right]&=&\frac{b^{\prime}(r)}{r^2}\label{re}
\\
8\pi p(r)\left[1-a(\rho(r)-3p(r))\right]&=&\frac{2f^{\prime}(r)}{r}\left(1-\frac{b(r)}{r}\right)-\frac{b(r)}{r^3}\nonumber
\\
\label{pe}
\end{eqnarray}

where $"\prime"$ denotes the derivative with respect to the radial coordinate $r$. $\rho(r)$, $p(r)$ are the matter  density and  pressure of the original matter configuration, respectively and  $a = \frac{\lambda}{8\pi}$ is non zero constant.\\
The effective density and pressure of the effective matter configuration are given by
\begin{eqnarray}
\rho_e(r)&=& \rho_{effective}(r)=8\pi \rho(r)\left[1-a(\rho(r)-3p(r))\right]\label{re1}
\\
p_e(r)&=& p_{effective}(r)=8\pi p(r)\left[1-a(\rho(r)-3p(r))\right]\label{pe1}
\end{eqnarray}
One can easily note that there is two independent field equations(\ref{re})-(\ref{pe}) of four unknowns $\rho(r)$, $p(r)$, $f(r)$ and $b(r)$ and hence it is difficult to find the exact expressions of $\rho(r)$, $p(r)$ in terms of the radial coordinate $r$ only. Therefore, in the next section, we will proceed to solve this system of equations accurately by assuming a linear equation of state(EoS) and  a redshift function.

\begin{figure}[h]
\centering
\includegraphics[width=6.5cm]{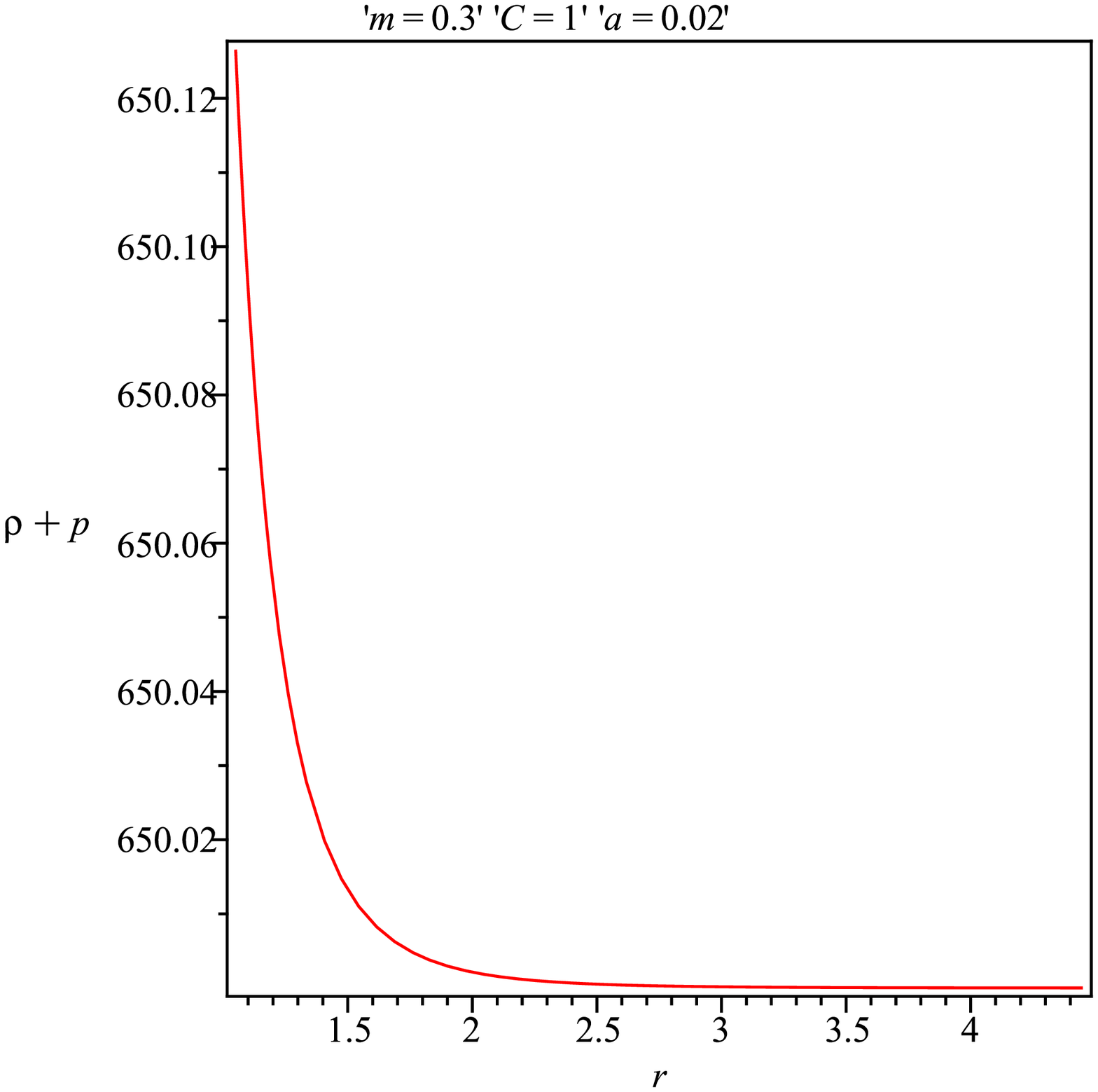}
\includegraphics[width=6.5cm]{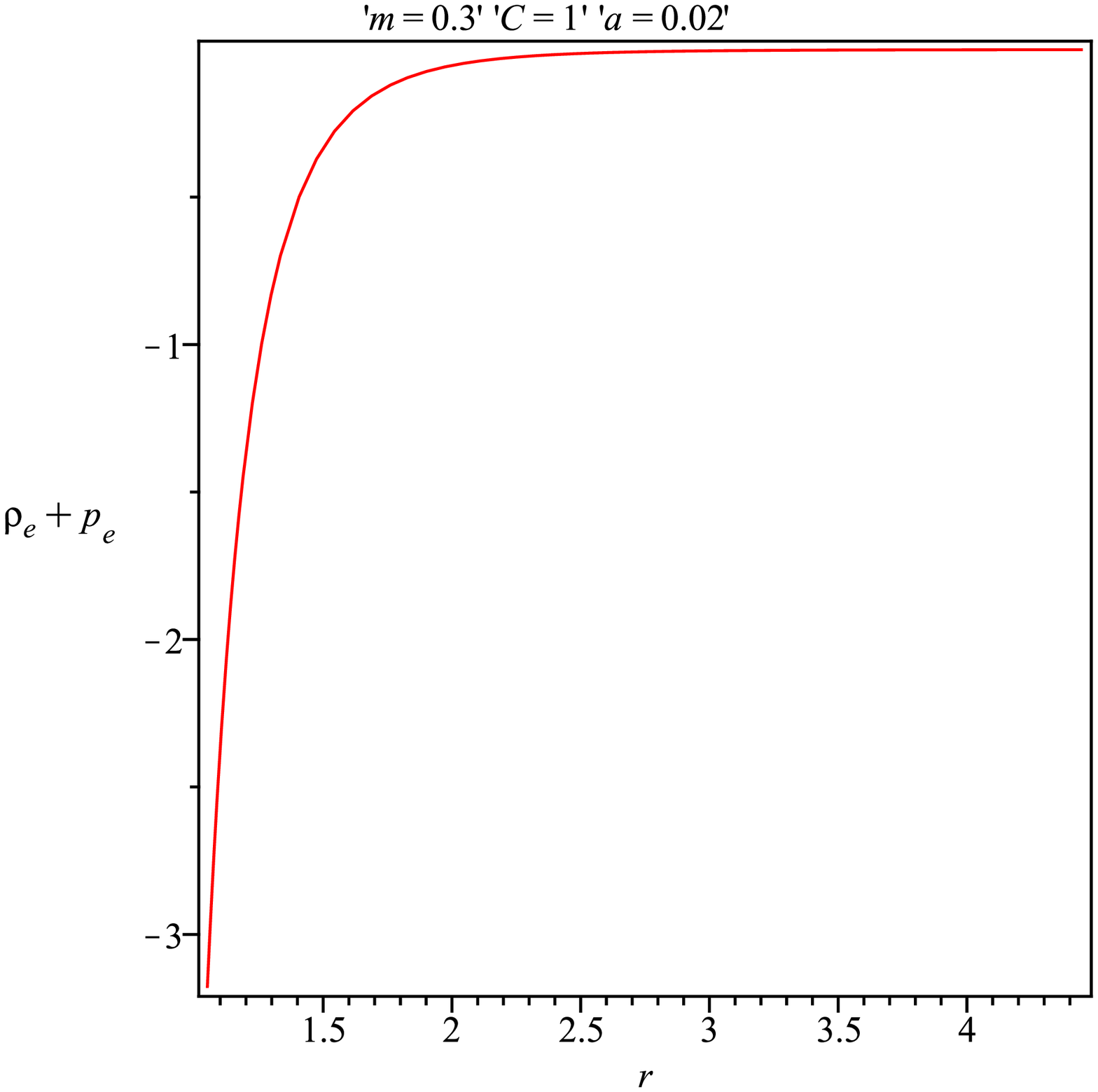}
\caption{The sum of original matter density and pressure(Left)  and  The sum of effective matter density and pressure(Right) are plotted  against  the radial coordinate $r ( >$ the throat radius) corresponding to $m = 0.3, C = 1, a = 0.02$.}\label{fig3}
\end{figure}

\section{Wormhole Solution}
To obtain the exact solutions of Einstein's field equation in the framework of $\kappa(R,T)$ gravity we can proceed  as follows :\\
 From Eqs. (\ref{re})- (\ref{pe})  we get
 \begin{equation}
\frac{\rho(r)}{p(r)}=\frac{\frac{b^{\prime}(r)}{r^2}}{\frac{2f^{\prime}(r)}{r}\left(1-\frac{b(r)}{r}\right)-\frac{b(r)}{r^3}}\label{rp}
\end{equation}
 Now, we consider a redshift function $f(r) = A$(a constant), where $A$ being a constant  avoids the event horizon and linear equation of state(EoS) $p(r) = m\rho(r)$, where m is a constant satisfying $0 < m < 1$  for the real matter configuration. Later we will provide a restriction on m as $ m \neq \frac{1}{3}$ for our solution.  For these EoS and redshift function, Eq.(\ref{rp}) yields
 \begin{eqnarray}
 b^\prime(r)+\frac{1}{mr}b(r)&=& 0\label{diff}
 \end{eqnarray}
 The differential Eq.(\ref{diff}) gives the following  shape function
 \begin{equation}
b(r)=Cr^{-\frac{1}{m}}\label{b}
\end{equation}
where C is an integration constant.

From the equation $b(r_0) = r_0$ we get the throat radius of the wormhole as $r_0 = C^{\frac{m}{m+1}}$. For the choice of $C = 1,$ the throat radius of the wormhole is $r_0 = 1$. Also, we provide some graphical representations  to get an overview of the obtained shape function as well as the conditions that need to satisfy for the  existence of wormhole structure. Fig.\ref{fig1} shows  the positive and decreasing behavior of the shape function with respect to the radial coordinate $r$(Left) and the right plot of Fig.\ref{fig1} presents the behavior of the ratio function $\frac{b(r)}{r}$, which shows that $\frac{b(r)}{r}$ tends zero for large values of the radial coordinate $r$ and hence the shape function satisfies the asymptotic behavior. Also, the shape function satisfies the flare out condition, (see Fig.\ref{fig2} (Left)). Consequently, our obtained shape function is well-behaved to present a wormhole structure.

Therefor, we obtain a wormhole representing line element as:

 \begin{equation}
ds^2 = e^{2A}dt^2-\left(1-Cr^{-\frac{m+1}{m}}\right)^{-1}dr^2-r^2(d\theta^2 + sin^2\theta d\phi^2)\label{wm}
\end{equation}

Now, we obtain the expression of  original matter density from Eq. (\ref{re}) using Eq.(\ref{b}) and the EoS as:

\begin{eqnarray}
\rho(r)&=& \frac{1}{2a(1-3m)}\left[1\pm \sqrt{1+\frac{aC(1-3m)}{2\pi m}r^{-\left(\frac{3m+1}{m}\right)}}\right]
\end{eqnarray}
with $m \neq \frac{1}{3}$.\\

\begin{figure}[h]
\centering
\includegraphics[width=6.5cm]{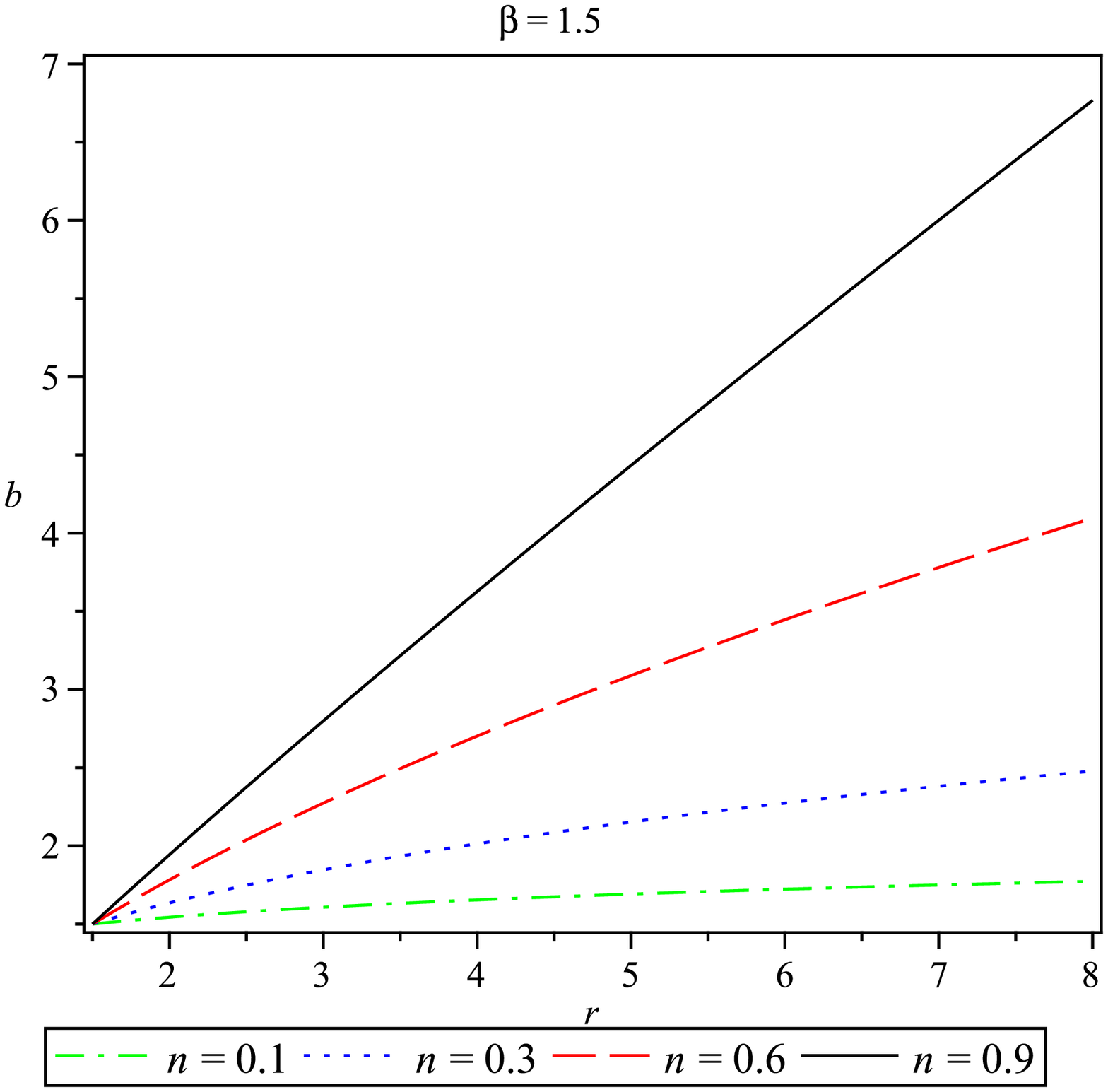}
\includegraphics[width=6.5cm]{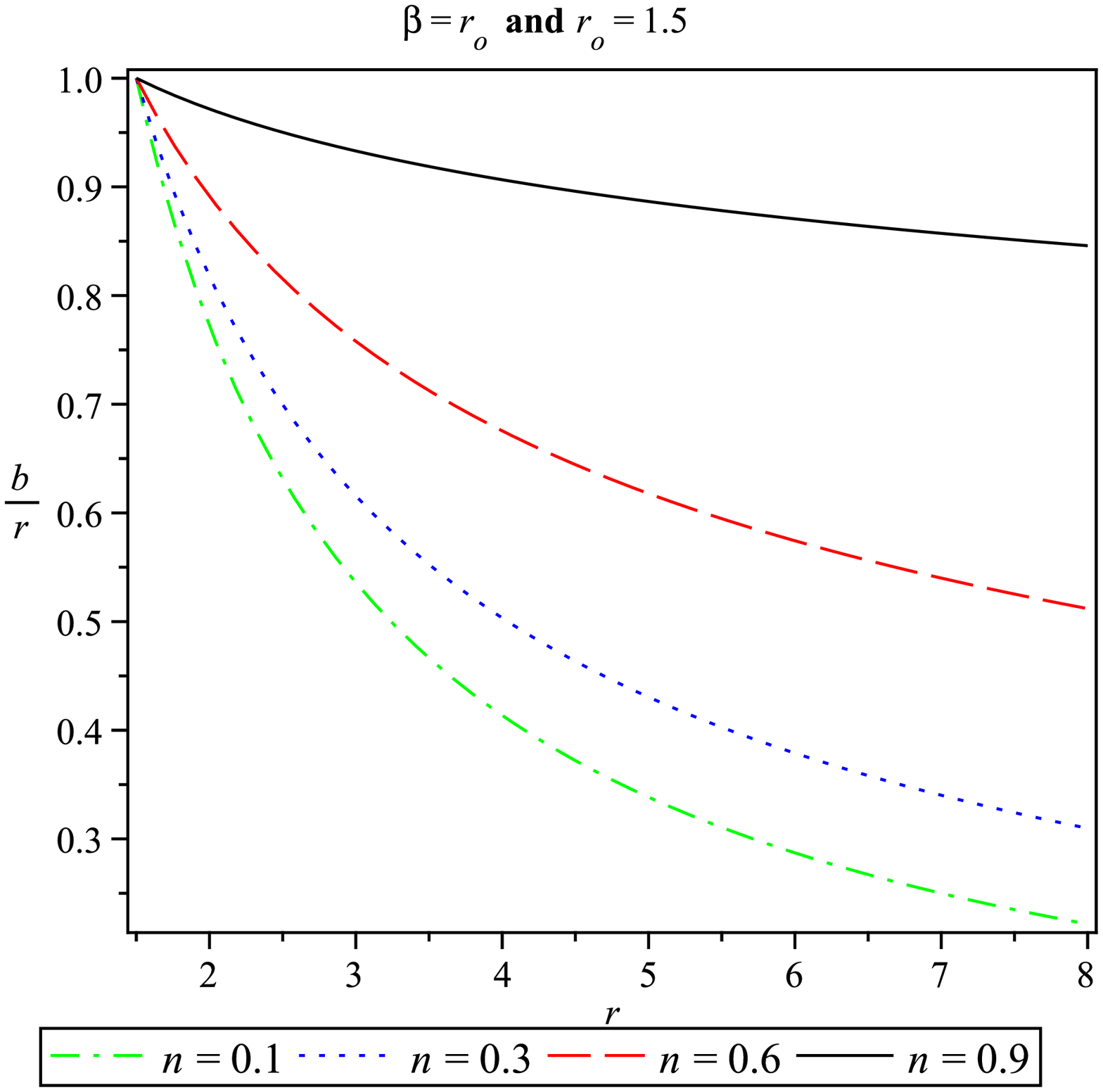}
\caption{ The shape function(Left) and $\frac{b(r)}{r}$(Right) are plotted  against  the radial coordinate $r ( \geq $ the throat radius) corresponding to $\beta = r_0 = 1.5$ and $n = 0.1, 0.3, 0.6, 0.9$.}\label{fig4}
\end{figure}

For further study, we consider  the following expression of $\rho(r)$:
\begin{eqnarray}
\rho(r)&=& \frac{1}{2a(1-3m)}\left[1+ \sqrt{1+\frac{aC(1-3m)}{2\pi m}r^{-\left(\frac{3m+1}{m}\right)}}\right]\label{rr}
\end{eqnarray}
Actually, we are interested to find the original matter as the real feasible matter configuration. For this purpose, we have taken one expression of $\rho(r)$ given in above Eq.(\ref{rr}). The original matter configuration becomes the exotic matter for another expression of $\rho(r)$.\\
Therefore, we obtain the following expressions
 \begin{equation}
\rho(r)+p(r)=\frac{(1+m)}{2a(1-3m)}\left[1+ \sqrt{1+\frac{aC(1-3m)}{2\pi m}r^{-\left(\frac{3m+1}{m}\right)}}\right]
\end{equation}

and
\begin{eqnarray}
\rho_{e}(r)+p_{e}(r)&=& \frac{8\pi(1+m)}{2a(1-3m)}\left[1+ \sqrt{1+\frac{aC(1-3m)}{2\pi m}r^{-\left(\frac{3m+1}{m}\right)}}\right]\left[1-\frac{1}{2}\left\{1+ \sqrt{1+\frac{aC(1-3m)}{2\pi m}r^{-\left(\frac{3m+1}{m}\right)}}\right\}\right]\nonumber
\\
\end{eqnarray}

It is well-known that the exotic matter is necessary to hold a wormhole structure and the exotic matter is characterized  by the violation of null energy condition(NEC). The energy conditions are defined as:
\begin{itemize}
 \item[(i)] The weak energy condition(WEC):  $T_{\alpha \beta}U^\alpha U^\beta\geq0$ i.e. $\rho(r)\geq0$ and $\rho(r)+p(r)\geq0$, where $T_{\alpha\beta}$ is the energy momentum tensor with $U^\alpha$ being a timelike vector.\\
\item[(ii)] The null energy condition(NEC):  $T_{\alpha\beta}\upsilon^\alpha \upsilon^\beta\geq0$  i.e.  $\rho(r)+p(r)\geq0$ and for the modified matter configuration $\rho_{e}(r)+p_{e}(r) \geq0 $, with $\upsilon^\alpha$ being a null vector.
\end{itemize}

  The density $\rho(r)$ of the original matter configuration is positive against the radial coordinate $r$ (see Fig.\ref{fig2}~(Right)). The original matter distribution satisfies the week energy condition(WEC) ($\rho(r) \geq 0, ~\rho(r) + p(r) \geq 0$) as well as the null energy condition(NEC) ($\rho(r) + p(r) \geq 0$) (see Fig.\ref{fig2}(Right) and Fig.\ref{fig3}(Left)) and hence the original matter distribution is of real feasible matter while the right plot of Fig.\ref{fig3} shows that  $\rho_{e}(r) + p_{e}(r)$ is negative i.e. the modified matter  violates  the null energy condition(NEC). Thus, we obtain an interesting result that a real feasible matter distribution will provide the fuel to construct and sustain a wormhole in $\kappa(R,T)$ gravity.

\begin{figure}[h]
\centering
\includegraphics[width=6.5cm]{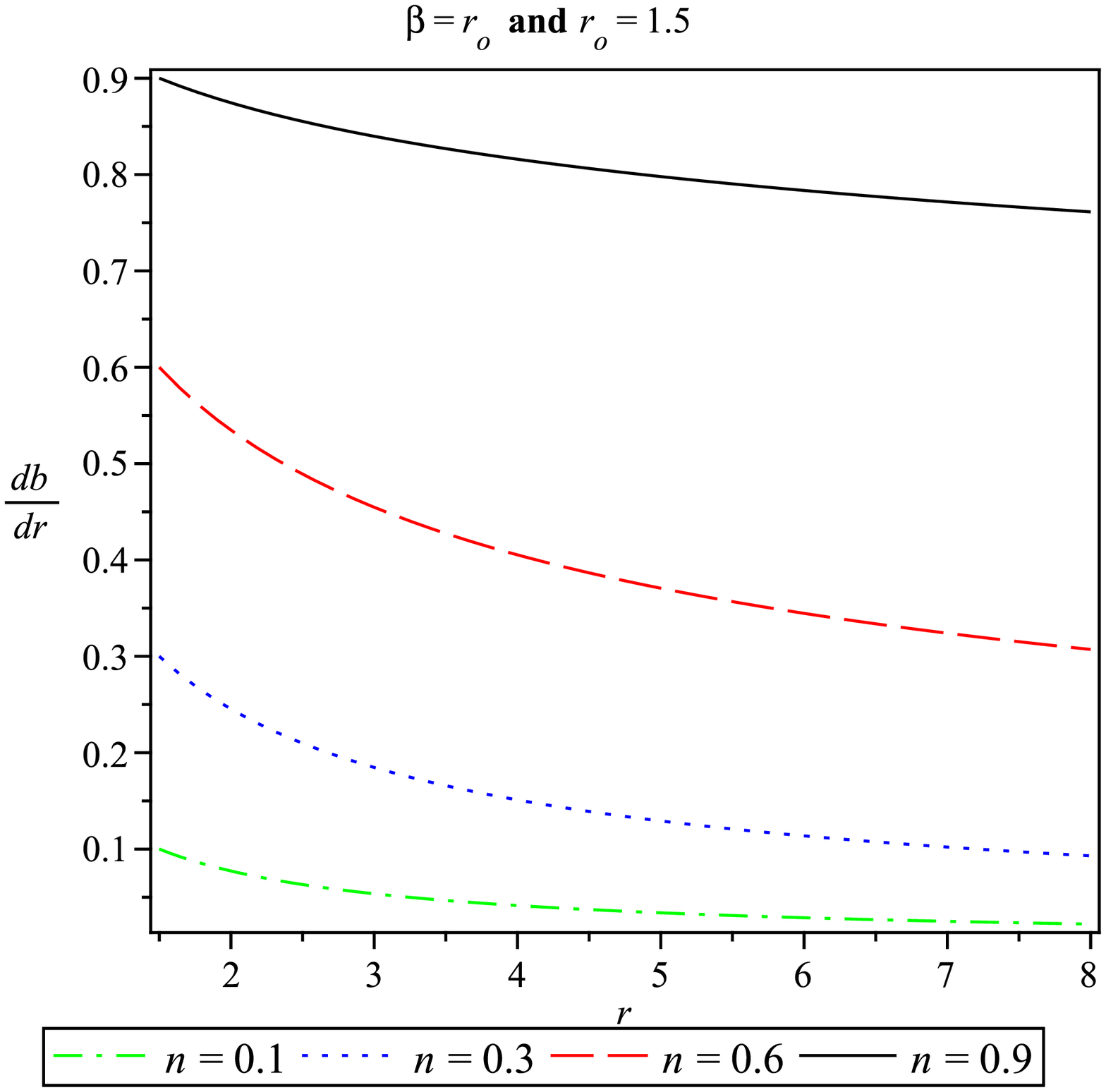}
\includegraphics[width=6.5cm]{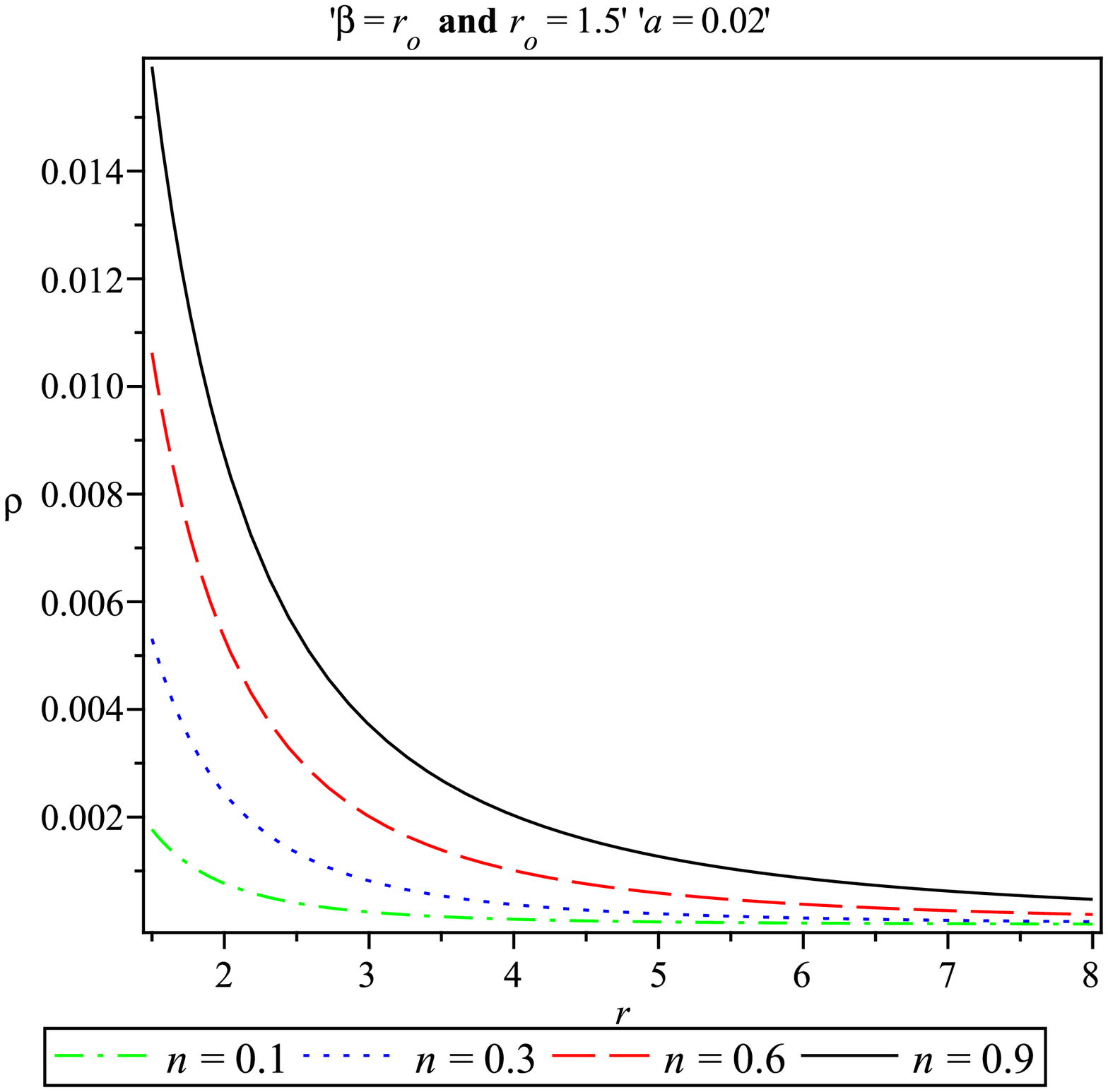}
\caption{  $\frac{db(r)}{dr}$(Left) and the original matter density(Right) are plotted  against  the radial coordinate $r ( \geq $ the throat radius) corresponding to $\beta = r_0 = 1.5$, $n = 0.1, 0.3, 0.6, 0.9$ and $\beta = r_0 = 1.5$, a = 0.02, $n = 0.1, 0.3, 0.6, 0.9$, respectively.}\label{fig5}
\end{figure}

\section{Wormhole solution with specific redshift and shape functions}

In this section, we will examine the existence of the wormhole structures by considering four pairs of two redshift functions and two shape functions into the account. One of the most important factor to measure the amount of exotic matter need to form a wormhole is "volume integral quantifier", which is  defined as:
\begin{equation}
I_v=\int\{\rho(r)+p(r)\}dV\label{vq}
\end{equation}
where $dV$ is the volume element, given by $dV = r^2\sin\theta dr d\theta$. One can note that  the volume integral quantifier $I_v~\rightarrow $ 0 if  $\rho(r) + p(r) \rightarrow 0$ and this reveals that  an arbitrarily small amount of exotic matter is responsible to construct a wormhole \cite{fr06}. Using this property later we will  show that an infinitesimal exotic matter is sufficient to construct the wormholes. \\
Now,  Eqs. (\ref{re}) and (\ref{pe}) are  the quadratic equations in $\rho(r)$ and $p(r)$, respectively. So, these equations give two different expressions for $\rho(r)$ and $p(r)$. After solving Eqs.(\ref{re})-(\ref{pe}) we obtain the general expressions of $\rho(r)$ and $p(r)$ in term of the redshift and shape functions as:

\begin{eqnarray}
\rho(r)&=&\frac{b^{\prime}(r)\left[2\pi r^2b(r)(1+2rf^{\prime}(r))-4\pi r^4f^{\prime}(r)\pm \sqrt{2\pi r(b(r)-2r^2f^{\prime}(r)+2b(r)rf^{\prime}(r))^2f_1}\right]}{4\pi ar  [r(b^{\prime}(r)-6rf^{\prime}(r))+b(r)(3+6rf^{\prime}(r))][b(r)-2r^2f^{\prime}(r)+2rbf^{\prime}(r)]}\nonumber
\\\label{rho1}
\\
p(r)&=&\frac{4\pi r^4f^{\prime}(r)-2\pi r^2b(r)(1+2rf^{\prime}(r))\mp \sqrt{2\pi r(b(r)-2r^2f^{\prime}(r)+2b(r)rf^{\prime}(r))^2f_1}}{4\pi ar^2  [r(b^{\prime}(r)-6rf^{\prime}(r))+b(r)(3+6rf^{\prime}(r))]}\label{p1}
\end{eqnarray}

whereas
\begin{eqnarray}
f_1&=&r(2\pi r^2-ab^{\prime}(r)+6arf^{\prime}(r))-3b(a+2arf^{\prime}(r))\nonumber
\end{eqnarray}

Therefore, anyone can easily  discuss the nature of the matter configuration by finding the  density $\rho(r)$ and pressure $p(r)$ from Eqs.(\ref{rho1})-(\ref{p1}) for the arbitrary choice of redshift and shape functions in $\kappa(R,T)$ gravity.\\
For further study we consider

\begin{eqnarray}
\rho(r)&=&\frac{b^{\prime}(r)\left[2\pi r^2b(r)(1+2rf^{\prime}(r))-4\pi r^4f^{\prime}(r)- \sqrt{2\pi r(b(r)-2r^2f^{\prime}(r)+2b(r)rf^{\prime}(r))^2f_1}\right]}{4\pi ar  [r(b^{\prime}(r)-6rf^{\prime}(r))+b(r)(3+6rf^{\prime}(r))][b(r)-2r^2f^{\prime}(r)+2rbf^{\prime}(r)]}\nonumber
\\\label{rho1}
\\
p(r)&=&\frac{4\pi r^4f^{\prime}(r)-2\pi r^2b(r)(1+2rf^{\prime}(r))+ \sqrt{2\pi r(b(r)-2r^2f^{\prime}(r)+2b(r)rf^{\prime}(r))^2f_1}}{4\pi ar^2  [r(b^{\prime}(r)-6rf^{\prime}(r))+b(r)(3+6rf^{\prime}(r))]}\label{p1}
\end{eqnarray}

because the other expressions are  not well-behaved for our models.

\begin{figure}[h]
\centering
\includegraphics[width=6.5cm]{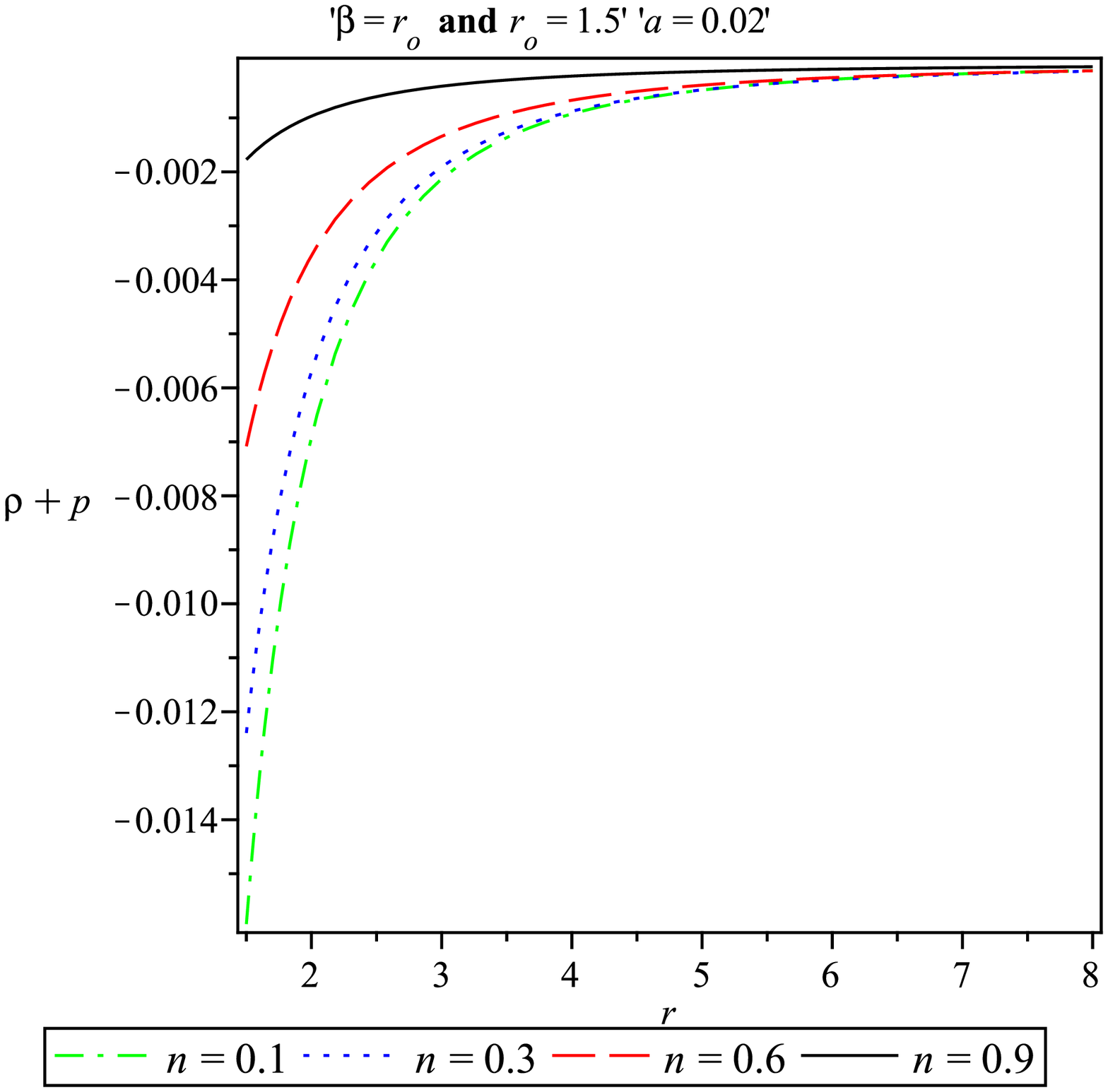}
\includegraphics[width=6.5cm]{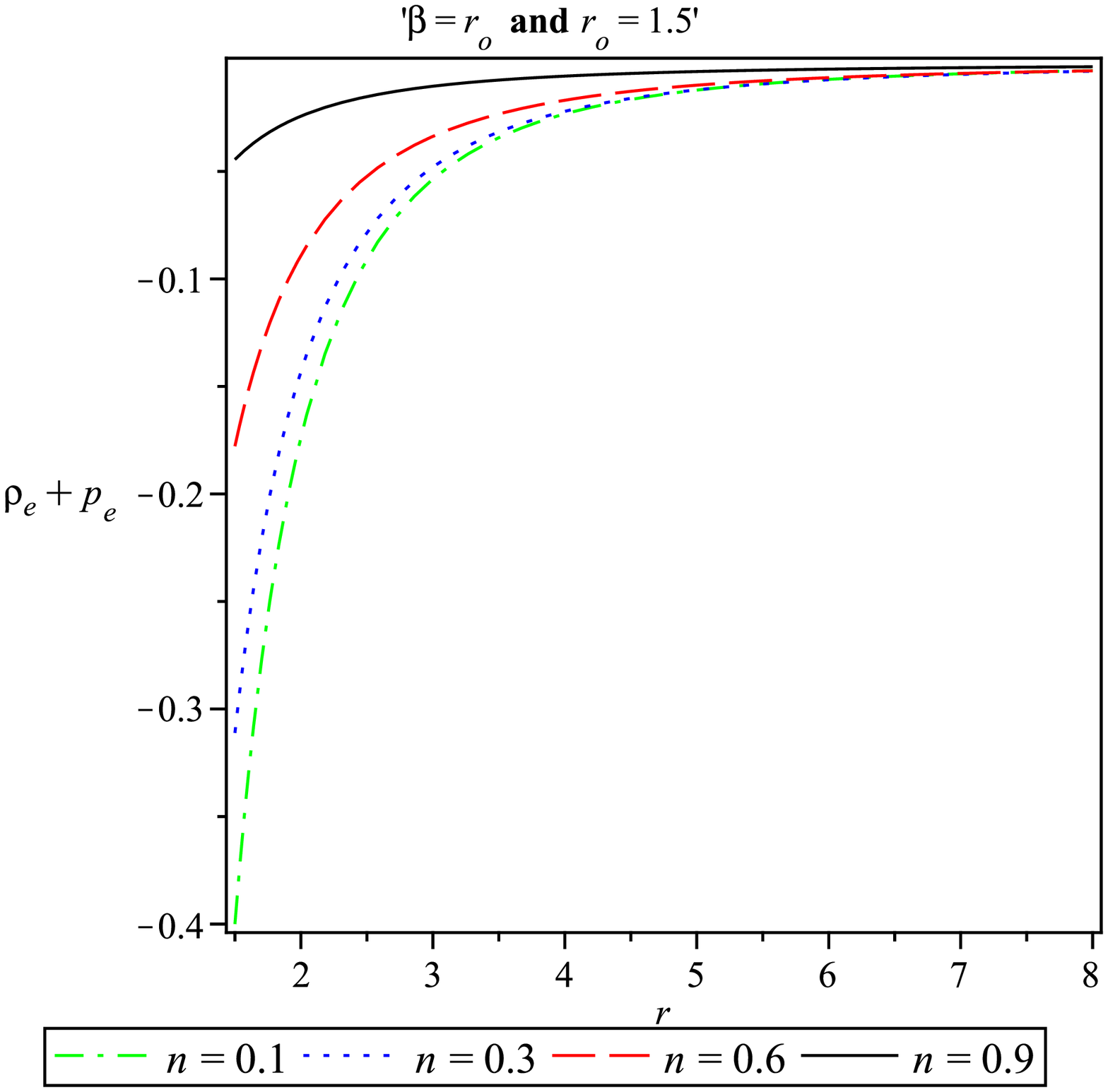}
\caption{ The sum of original matter density and pressure(Left) and  the sum of effective matter density and pressure(Right) are plotted  against  the radial coordinate $r ( \geq $ the throat radius) corresponding to $\beta = r_0 = 1.5$, $ a = 0.02$, $n = 0.1, 0.3, 0.6, 0.9$ and $\beta = r_0 = 1.5$, $n = 0.1, 0.3, 0.6, 0.9$, respectively.}\label{fig6}
\end{figure}

\subsection{ The shape function $ b(r)=\beta \left(\frac{r}{\beta}\right)^n $  }

Here, we consider the shape function $ b(r)=\beta \left(\frac{r}{\beta}\right)^n $\cite{fr19} to analyze the solutions with respect to two different redshift functions, separately. The throat radius of the wormhole is  $r_0 = \beta$ where $\beta \left(\frac{r_0}{\beta}\right)^n = r_0$. The shape function is positive and increasing in nature, shown in Fig.\ref{fig4}(Left).  The Fig.\ref{fig4}(Right) ensures that $\frac{b(r)}{r}$ tends to zero for the large values of the radial coordinate $r$ i.e. the shape function has asymptotic behavior.  It is obvious from Fig.\ref{fig5}(Left) that ~$b^\prime (r \geq r_0) < 1$ for $n < 1$ i.e. the shape function satisfies the flare-out condition for $n < 1$.\\
Next, we will discuss the solutions for the above shape function along with two different redshift functions, separately.

\begin{figure}[h]
\centering
\includegraphics[width=6.5cm]{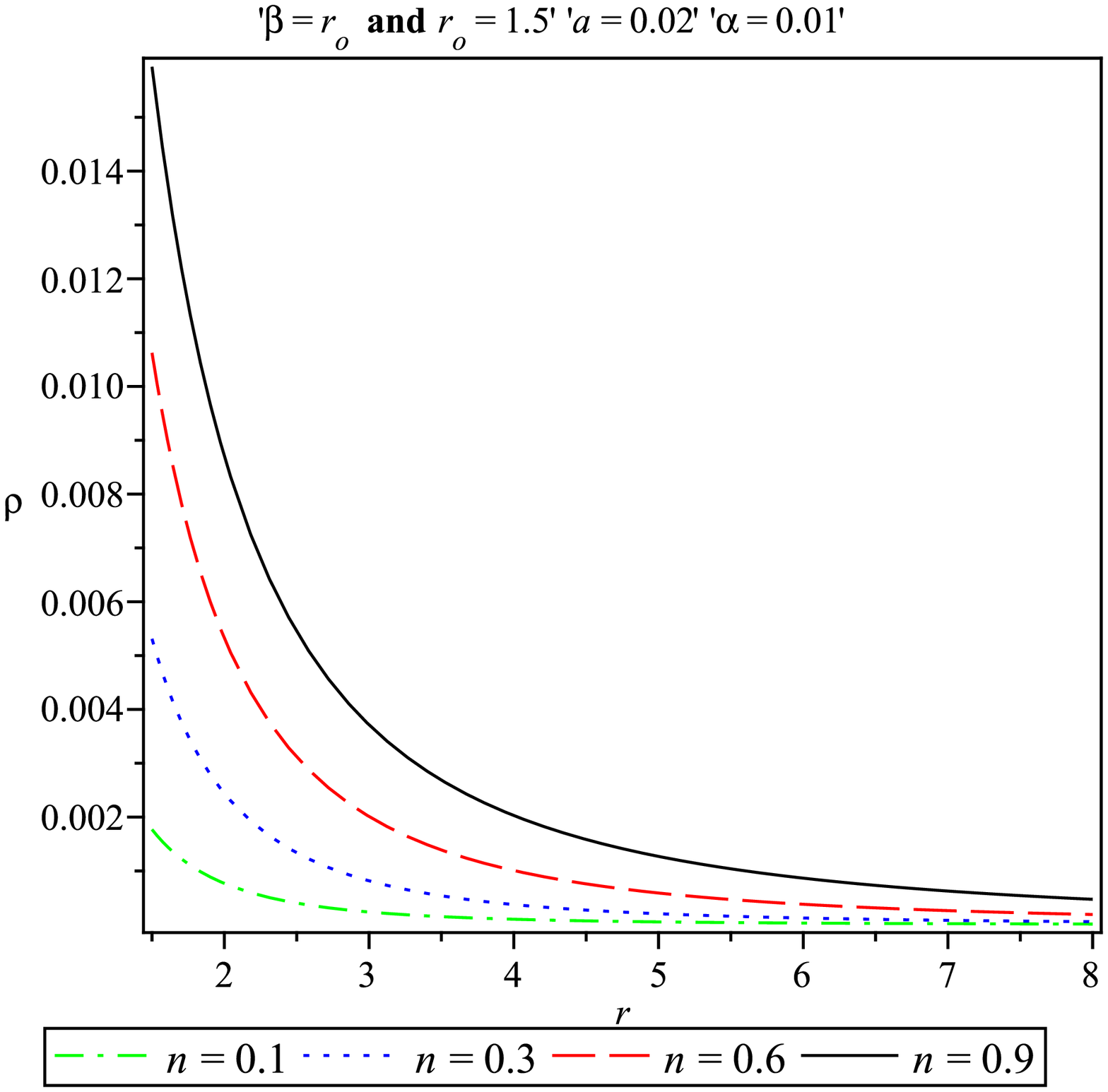}
\includegraphics[width=6.5cm]{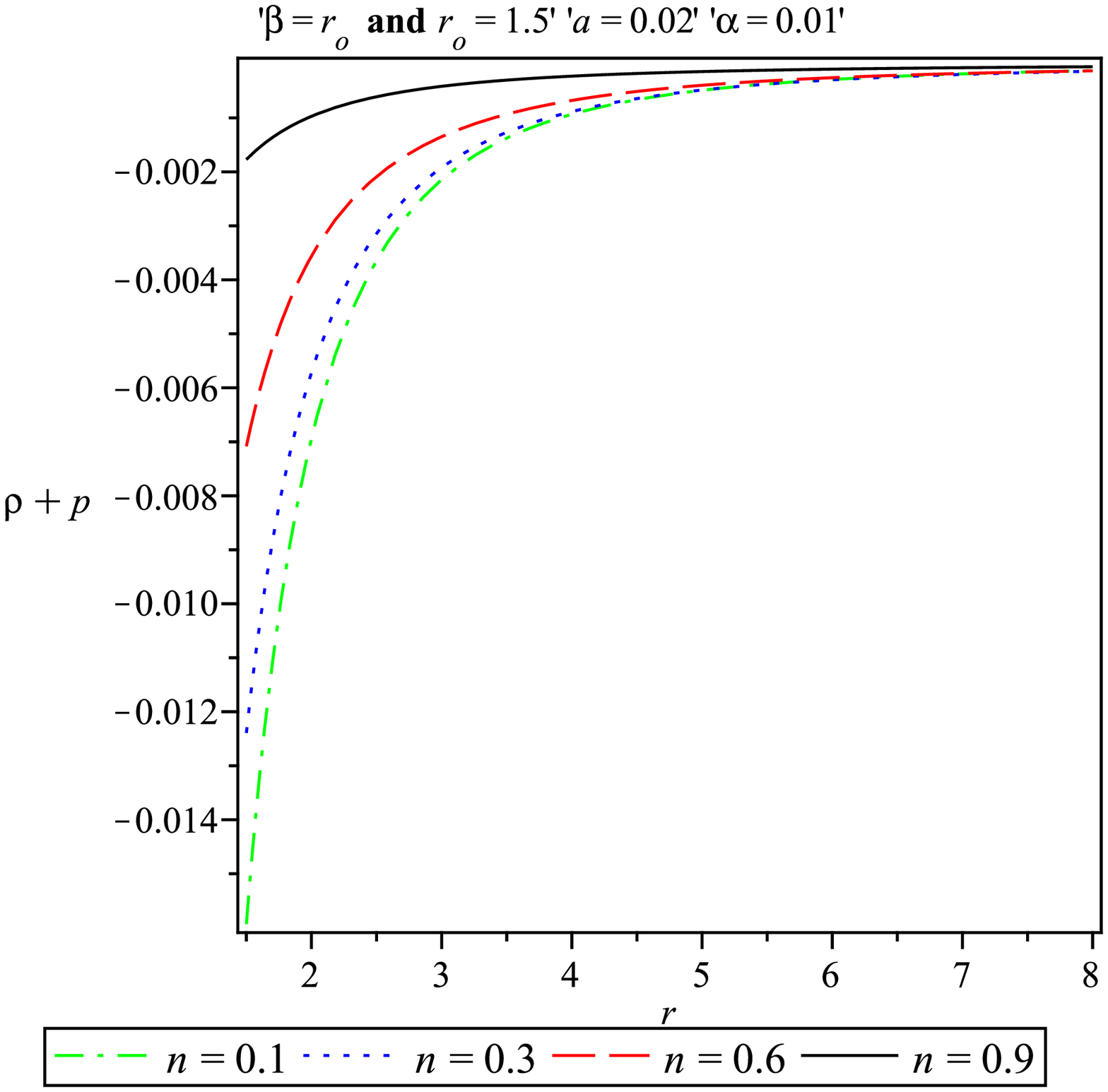}
\caption{The original matter density(Left) and   The sum of original matter density and pressure(Right) are plotted  against the radial coordinate $r ( \geq $ the throat radius) corresponding to $\beta = r_0 = 1.5$, $ a = 0.02$, $\alpha = 0.01$ and $n = 0.1, 0.3, 0.6, 0.9$.}\label{fig7}
\end{figure}

\subsubsection{ The redshift function $f(r)=0$}
The constant redshift function $f(r)$, known as the tidal force solution is always admissible as the constant redshift function is finite and avoids the event horizon.

For the tidal force solution and above shape function,  Eqs.(\ref{rho1})-(\ref{p1}) yields the expressions of original matter density $\rho(r)$ and pressure $p(r)$ as:

\begin{eqnarray}
\rho(r)&=&\frac{n}{4\pi ar^2(3+n)}\left[2\pi r^2- \sqrt{2\pi r\left\{2\pi r^3-a(3+n)\beta \left(\frac{r}{\beta}\right)^n\right\}}\right]\label{rho12}
\\
p(r)&=&-\frac{1}{4\pi ar^2(3+n)}\left[2\pi r^2- \sqrt{2\pi r\left\{2\pi r^3-a(3+n)\beta \left(\frac{r}{\beta}\right)^n\right\}}\right]\label{p12}
\end{eqnarray}
Therefore,
\begin{eqnarray}
\rho(r) + p(r)&=&\frac{n-1}{4\pi ar^2(3+n)}\left[2\pi r^2- \sqrt{2\pi r\left\{2\pi r^3-a(3+n)\beta \left(\frac{r}{\beta}\right)^n\right\}}\right]\label{rhop12}
\end{eqnarray}

Also, from Eqs.(\ref{re1}) and (\ref{pe1}) we obtain

\begin{eqnarray}
\rho_{e}(r)&=&\frac{n\beta\left(\frac{r}{\beta}\right)^n}{r^3}\label{rhoe11}
\\
p_{e}(r)&=&-\frac{\beta \left(\frac{r}{\beta}\right)^n}{r^3}\label{pe11}
\end{eqnarray}
and hence
\begin{eqnarray}
\rho_{e}(r)+p_{e}(r)&=&(n-1)\frac{\beta\left(\frac{r}{\beta}\right)^n}{r^3}\label{rhope11}
\end{eqnarray}

To check  the behavior of the obtained original matter density $\rho(r)$ and the null energy condition(NEC) we draw the graphs for $\rho(r)$ and $\rho(r) + p(r)$ and $\rho_{e}(r) + p_{e}(r)$ against the radial coordinate $r$   in Figs.\ref{fig5}(Right)-\ref{fig6}. The right sketch of Fig.\ref{fig5} shows the positive behavior of the original matter density with respect to the radial coordinate $r$ and  Fig.\ref{fig6} indicates that $\rho(r) + p(r) < 0$ and $\rho_{e}(r) + p_{e}(r) < 0$ i.e. the original matter distribution as well as the modified matter distribution in  $\kappa(R,T)$ gravity both violate the null energy condition(NEC) and hence there exists a wormhole structure supported by the matter configuration in $\kappa(R,T)$ gravity. Moreover, from Eq.(\ref{rhop12}) it is clear that $\rho(r) + p(r)$ tends to zero when $n$ tends to unity i.e. the volume integral quantifier~$I_v$ tends to zero  when $n$ tends to unity. So, an infinitesimal amount of exotic matter is sufficient to hold a wormhole.

\begin{figure}[h]
\centering
\includegraphics[width=6.5cm]{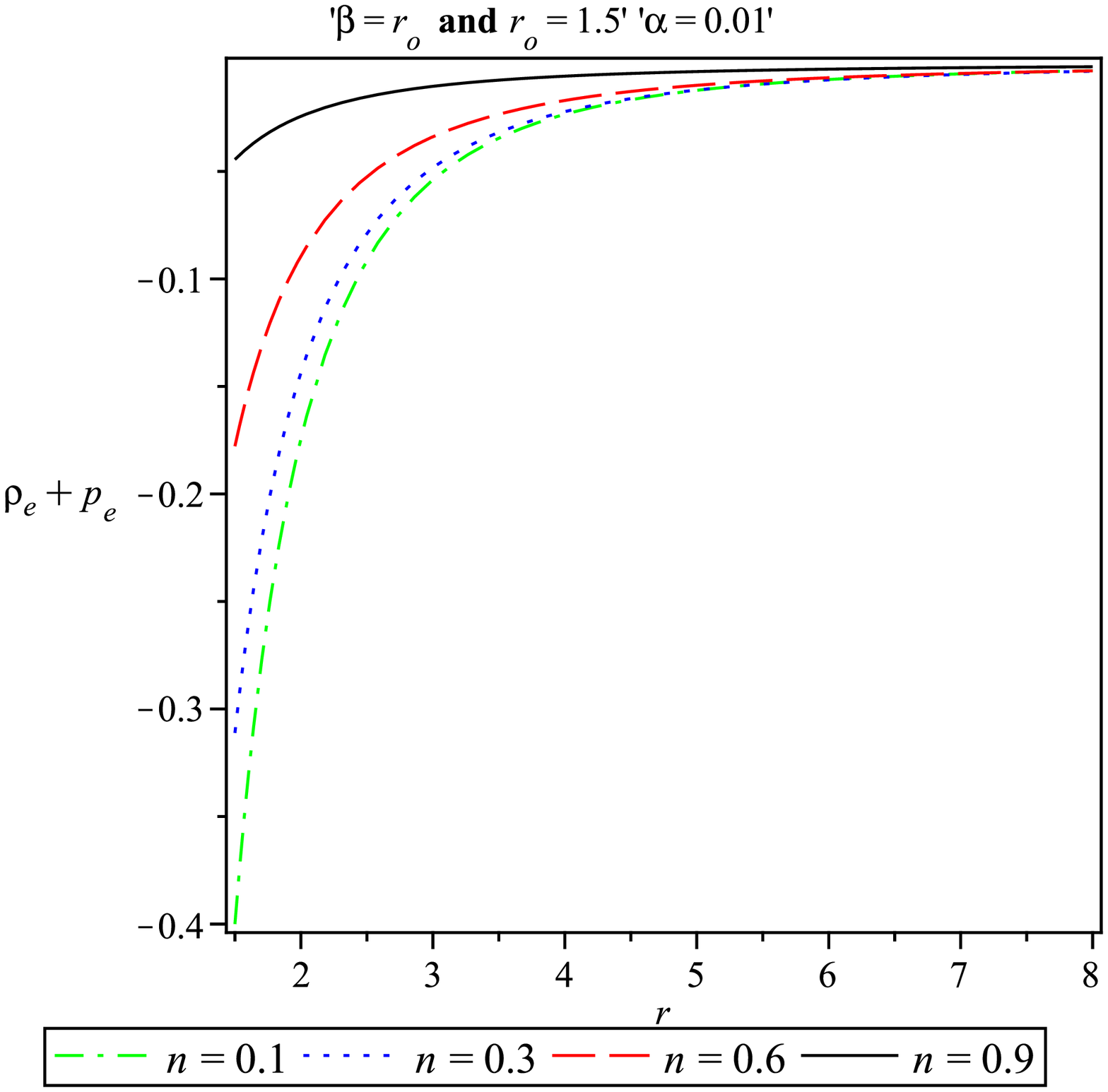}
\includegraphics[width=6.5cm]{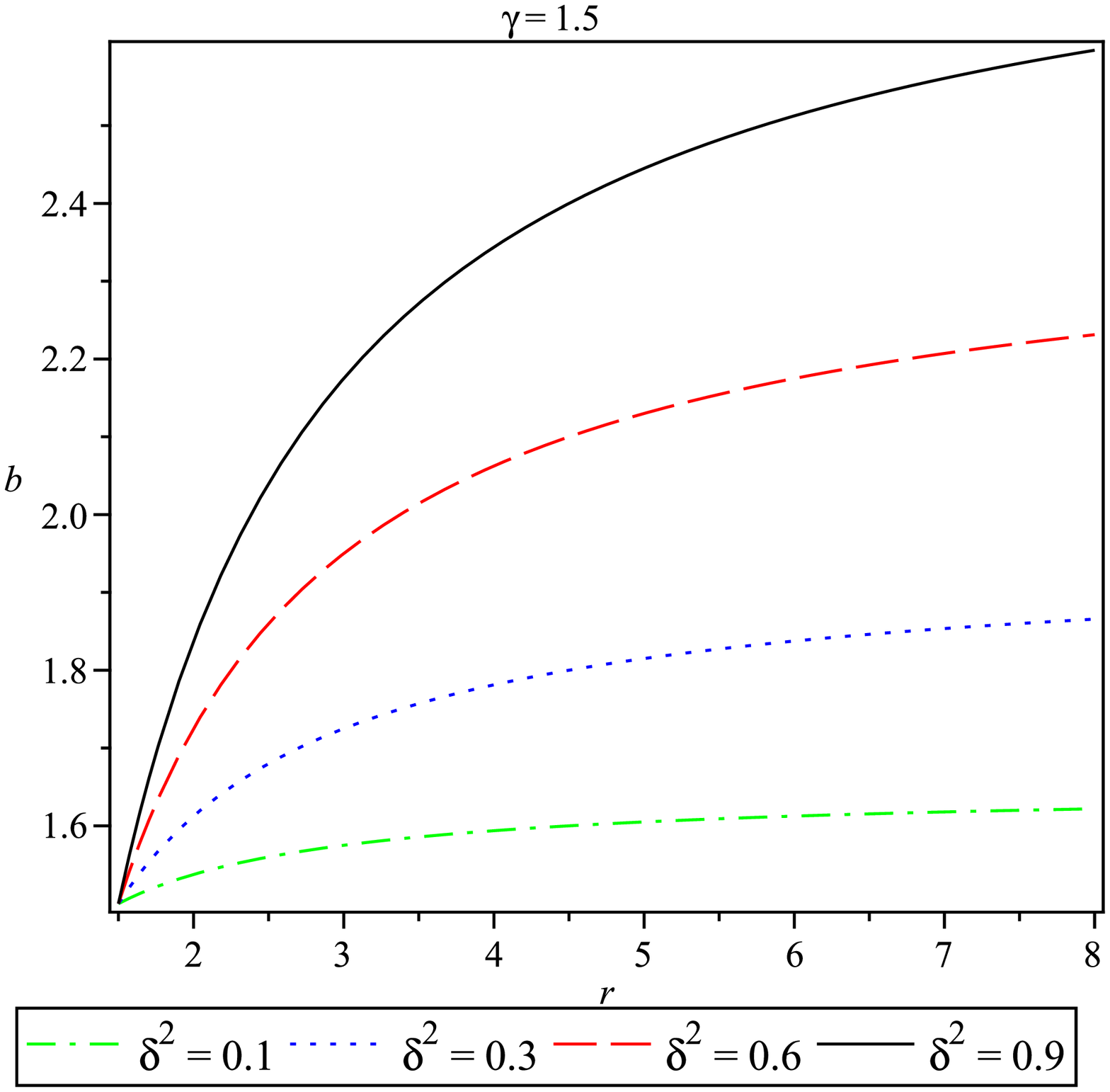}
\caption{The sum of effective matter density and pressure(Left) and  the shape function(Right) are plotted  against  the radial coordinate $r ( \geq $ the throat radius) corresponding to $\beta = r_0 = 1.5$, $\alpha = 0.01$, $n = 0.1, 0.3, 0.6, 0.9$ and $\gamma = r_0 = 1.5$ and $\delta^2 = 0.1, 0.3, 0.6, 0.9$, respectively.}\label{fig8}
\end{figure}

\subsubsection{  The redshift function $f(r)=\frac{\alpha}{r}$.}

The shape function $f(r)=\frac{\alpha}{r}$ is always finite for $r \geq$ $r_0$, where $\alpha$ is constant and hence this redshift function avoids the event horizon. Therefore, from Eqs.(\ref{rho1})$-$(\ref{p1}) we get the expressions for $\rho(r)$ and $p(r)$  as:

\begin{eqnarray}
\rho(r)&=&n\beta\left(\frac{r}{\beta}\right)^n\left[2\pi r^3\left\{2\alpha+\beta\left(\frac{r}{\beta}\right)^n\right\}-4\pi \alpha r^2\beta\left(\frac{r}{\beta}\right)^n- f_4\right]\left[{4\pi ar \left\{2\alpha r+(r-2\alpha)\beta \left(\frac{r}{\beta}\right)^n\right\}f_5}\right]^{-1}
\label{rho32}
\\
p(r)&=&-\frac{1}{4\pi ar^2f_5}\left[2\pi r^3\left\{2\alpha+\beta\left(\frac{r}{\beta}\right)^n\right\}-4\pi \alpha r^2\beta\left(\frac{r}{\beta}\right)^n- f_4\right]\label{p32}
\end{eqnarray}

 Therefore,
 \begin{eqnarray}
\rho(r)+p(r)&=&\left[2\pi r^3\left\{2\alpha+\beta\left(\frac{r}{\beta}\right)^n\right\}-4\pi \alpha r^2\beta\left(\frac{r}{\beta}\right)^n- f_4\right]\left[rn\beta\left(\frac{r}{\beta}\right)^n-\left\{2\alpha r+(r-2\alpha)\beta \left(\frac{r}{\beta}\right)^n\right\}\right]\nonumber
\\
 \left[4\pi ar^2 \left\{2\alpha r+(r-2\alpha)\beta \left(\frac{r}{\beta}\right)^n\right\}f_5\right]^{-1}
\\\label{rhop32}
\end{eqnarray}
whereas

\begin{eqnarray}
f_4&=&\left\{2\alpha r+(r-2\alpha)\beta \left(\frac{r}{\beta}\right)^n\right\}\sqrt{2\pi(2\pi r^4-af_5)}\nonumber\\
f_5&=&r\left[6\alpha+(3+n)\beta \left(\frac{r}{\beta}\right)^n\right]-6\alpha\beta \left(\frac{r}{\beta}\right)^n
\end{eqnarray}

Also, from Eqs.(\ref{re1})$-$(\ref{pe1}) we get the following result
\begin{eqnarray}
\rho_{e}(r)&=&\frac{n\beta\left(\frac{r}{\beta}\right)^n}{r^3}\label{rhoe21}
\\
p_{e}(r)&=&-\frac{2\alpha r +(r-2\alpha)\beta\left(\frac{r}{\beta}\right)^n}{r^4}\label{pe21}
\end{eqnarray}

and hence
\begin{eqnarray}
\rho_{e}(r)+p_{e}(r)&=&\frac{[2\alpha+r(n-1)]\beta\left(\frac{r}{\beta}\right)^n-2\alpha r}{r^4}\nonumber
\\\label{rhope21}
\end{eqnarray}

Fig.$\ref{fig7}$(Left) shows the positive  and decreasing behavior of original matter density $\rho(r)$. The matter configuration in $\kappa(R,T)$ gravity violates the null energy condition(NEC), clear from Figs.$\ref{fig7}$(Right)-$\ref{fig8}$(Left) and these confirm that the solution supports a wormhole structure. Moreover, from  Eq.(\ref{rhop32}) we can see that $\rho(r) + p(r)$ tends to zero whenever $n$ tends to unity. Therefore,  in this case, also an arbitrarily small amount of exotic matter is responsible to construct a wormhole design.

\begin{figure}[h]
\centering
\includegraphics[width=6.5cm]{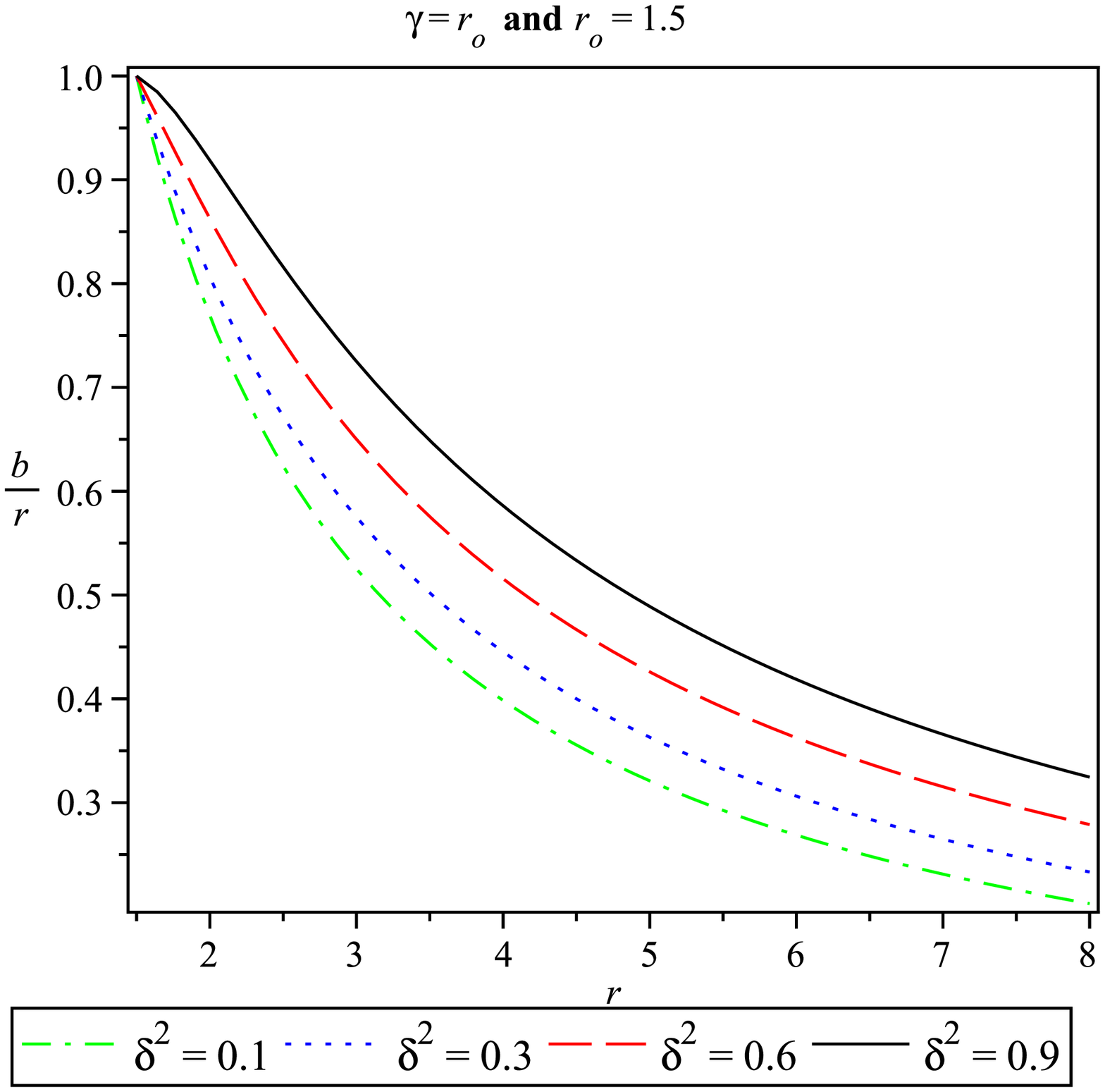}
\includegraphics[width=6.5cm]{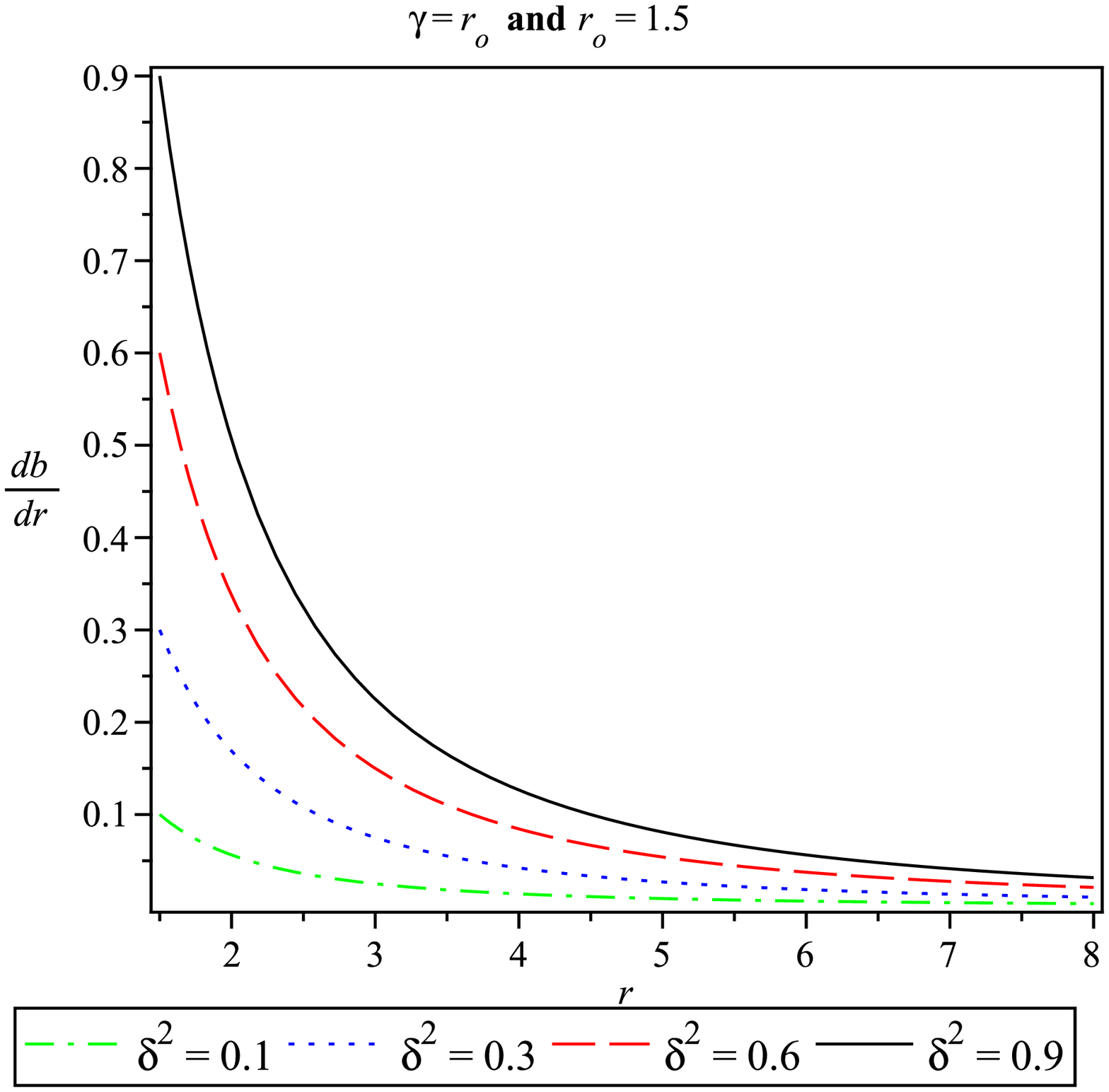}
\caption{ $\frac{b(r)}{r}$(Left) and  $\frac{db(r)}{dr}$(Right) are   plotted  against the radial coordinate $r ( \geq $ the throat radius) corresponding to $\gamma = r_0 = 1.5$ and $\delta^2 = 0.1, 0.3, 0.6, 0.9$.}\label{fig9}
\end{figure}

\subsection{The shape function $ b(r)=\gamma\left\{1+\delta^2\left(1-\frac{\gamma}{r}\right)\right\} $ .}

 we consider another form of the shape function $ b(r)=\gamma\left\{1+\delta^2\left(1-\frac{\gamma}{r}\right)\right\}$\cite{fr19} where $\gamma,  \delta $ are constants. The throat radius of the wormhole is obtained as $r_0 =  \gamma$ by using $\gamma\left\{1+\delta^2\left(1-\frac{\gamma}{r_0}\right)\right\} = r_0$. The shape function is positive and increasing in nature (see Fig.\ref{fig8}(Right)) and it has  asymptotic behavior (see Fig.\ref{fig9}(Left)). The shape function also satisfies the flare-out condition, shown in Fig.(\ref{fig9})(Right) provided $\delta^2 < 1$. \\
 Here, we take the same two  redshift functions $f(r) = 0$ and $f(r) = \frac{\alpha}{r}$  to analyze the corresponding solutions, separately.\\
Therefore,  in  similar way,  we obtain the following expressions for these sets of  redshift and shape functions:
\begin{eqnarray}
\rho(r)&=&\frac{2\pi \gamma \delta^2 r^2 + \gamma \delta^2 \sqrt{2\pi \left[2\pi r^4 -a\gamma \left\{3r(1+\delta^2)-2\gamma \delta^2\right\}\right]}}{4\pi ar^2 \left\{3r(1+\delta^2)-2\gamma \delta^2\right\}}\label{rho22}
\\
p(r)&=&-\frac{[2\pi \gamma \delta^2 r^2 + \gamma \delta^2 \sqrt{2\pi \left[2\pi r^4 -a\gamma \left\{3r(1+\delta^2)-2\gamma \delta^2\right\}\right]}]\{r+(r-\gamma)\delta^2\}}{4\pi ar^2 \gamma\delta^2 \left\{3r(1+\delta^2)-2\gamma \delta^2\right\}}\label{p22}
\\
\rho(r) +p(r)&=&\left\{\frac{2\pi \gamma \delta^2 r^2 + \gamma \delta^2 \sqrt{2\pi \left[2\pi r^4 -a\gamma \left\{3r(1+\delta^2)-2\gamma \delta^2\right\}\right]}}{4\pi ar^2 \left\{3r(1+\delta^2)-2\gamma \delta^2\right\}}\right\}\left\{1-\frac{\left[r+(r-\gamma)\delta^2\right]}{\gamma \delta^2 }\right\}\label{rhop22}
\end{eqnarray}

\begin{eqnarray}
\rho_{e}(r)&=&\frac{(\delta\gamma)^2}{r^4}\label{rhoe21}\label{rhoe12}
\\
p_{e}(r)&=&-\frac{\gamma\left\{(1+\delta^2)r-\gamma\delta^2\right\}}{r^4}\label{pe12}
\\
\rho_{e}(r)+p_{e}(r)&=&-\frac{\gamma\left\{(1+\delta^2)r-2\gamma\delta^2\right\}}{r^4}\label{rhope12}
\end{eqnarray}

\begin{figure}[h]
\centering
\includegraphics[width=6.5cm]{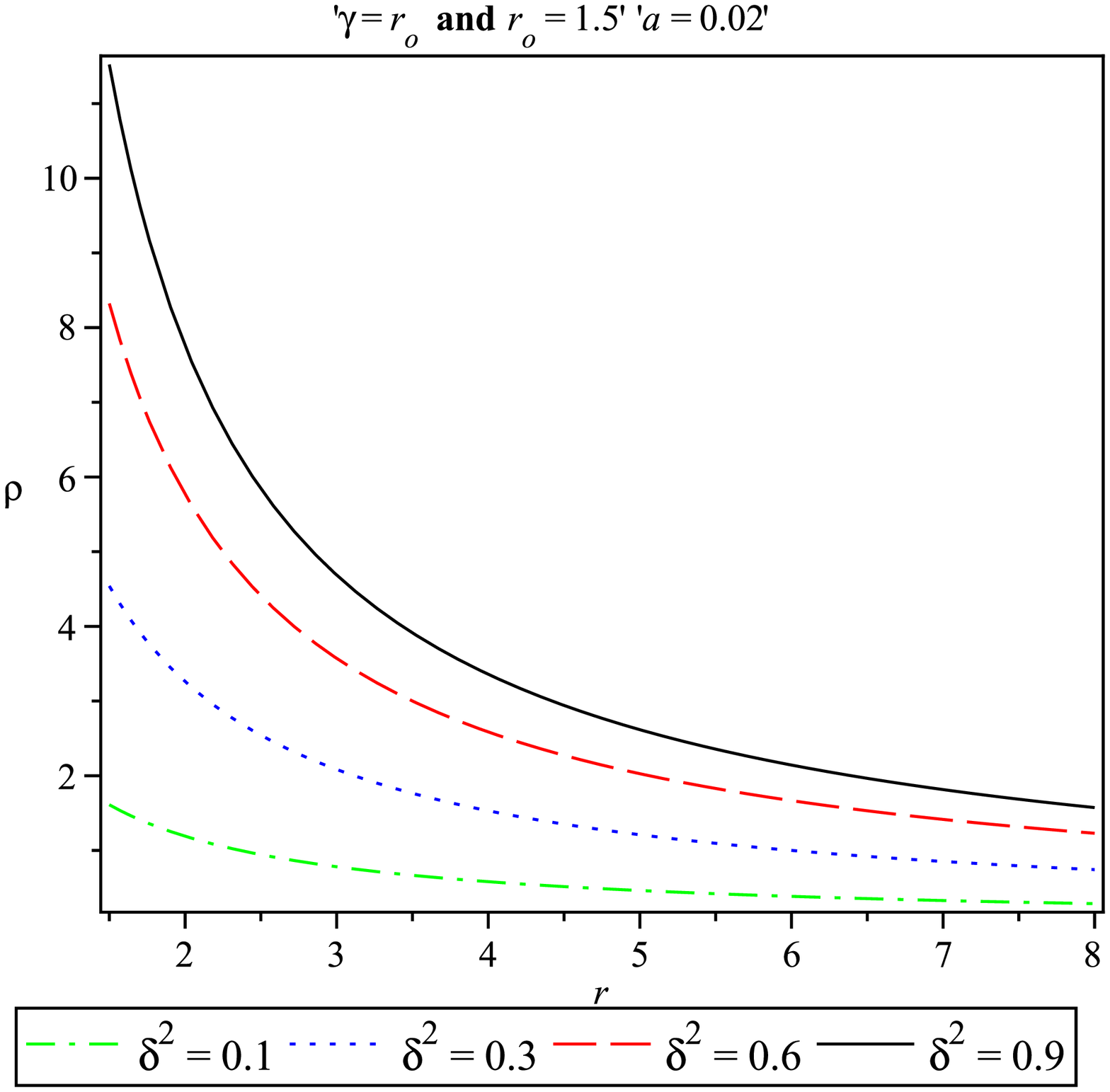}
\includegraphics[width=6.5cm]{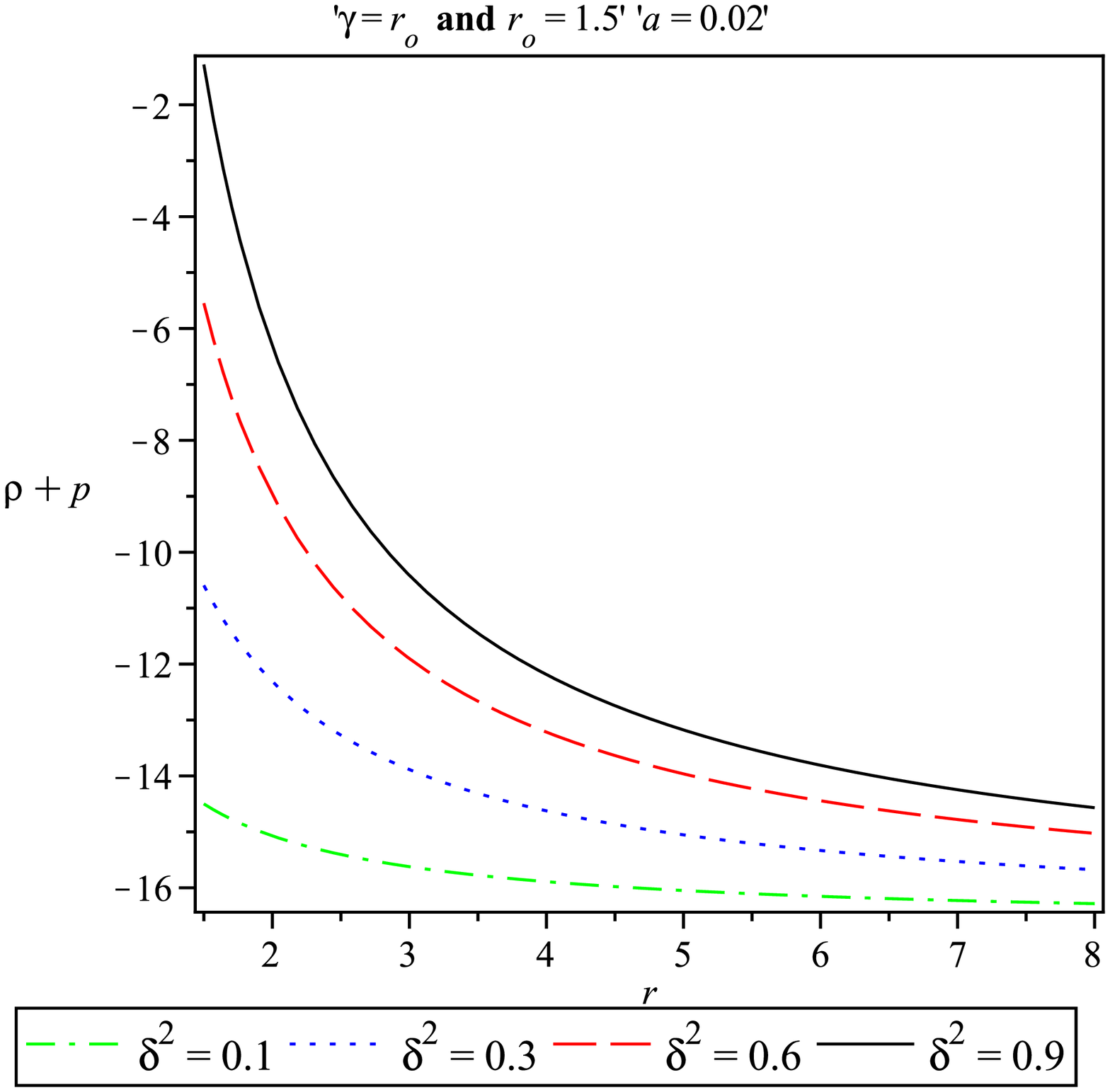}
\caption{The original matter matter density(Left) and the sum of original matter density and pressure(Right) are   plotted  against  the radial coordinate $r ( \geq $ the throat radius) corresponding to $\gamma = r_0 = 1.5$, $a = 0.02$ and $\delta^2 = 0.1, 0.3, 0.6, 0.9$.}\label{fig10}
\end{figure}

And
\begin{eqnarray}
\rho(r)&=&\frac{\gamma^2\delta^2\left[f_6 - \left\{2\alpha\gamma^2\delta^2+r^2(2\alpha+\gamma+\gamma\delta^2)-r\gamma(\gamma\delta^2+2\alpha+2\alpha\delta^2)\right\}\sqrt{2\pi r(2\pi r^5-a f_{7})}\right]}{4a\pi r^2\left\{2\alpha\gamma^2\delta^2+r^2(2\alpha+\gamma+\gamma\delta^2)-r\gamma(\gamma\delta^2+2\alpha+2\alpha\delta^2)\right\}f_{7}(r)}\label{rho41}\nonumber
\\
p(r)&=&-\frac{f_6 - \left\{2\alpha\gamma^2\delta^2+r^2(2\alpha+\gamma+\gamma\delta^2)-r\gamma(\gamma\delta^2+2\alpha+2\alpha\delta^2)\right\}\sqrt{2\pi r(2\pi r^5-a f_{7})}}{4a\pi r^3 f_7}\label{p14}
\\
\rho(r)+p(r)&=&\left\{\frac{f_6 - \left\{2\alpha\gamma^2\delta^2+r^2(2\alpha+\gamma+\gamma\delta^2)-r\gamma(\gamma\delta^2+2\alpha+2\alpha\delta^2)\right\}\sqrt{2\pi r(2\pi r^5-a f_7)}}{4a\pi r^2 f_7}\right\}\nonumber\\
&&\times\left\{\frac{\gamma^2\delta^2}{2\alpha\gamma^2\delta^2+r^2(2\alpha+\gamma+\gamma\delta^2)-r\gamma(\gamma\delta^2+2\alpha+2\alpha\delta^2)}-\frac{1}{r}\right\}\label{rhop14}
\end{eqnarray}

\begin{eqnarray}
\rho_{e}(r)&=&\frac{(\gamma\delta)^2}{r^4}\label{rhoe22}
\\
p_{e}(r)&=&\frac{r\gamma^2\delta^2-2\alpha \left\{r^2-\gamma (r+\delta^2-\gamma\delta^2)\right\}-r^2\gamma-r^2\gamma\delta^2}{r^5}\label{pe22}
\\
\rho_{e}(r)+p_{e}(r)&=&\frac{2r\gamma^2\delta^2-2\alpha \left\{r^2-\gamma (r+\delta^2-\gamma\delta^2)\right\}-r^2\gamma-r^2\gamma\delta^2}{r^5}\nonumber
\\\label{rhope22}
\end{eqnarray}
whereas
\begin{eqnarray}
f_6&=&2\pi r^3(2r^2\alpha+r^2\gamma-2r\alpha\gamma+r^2\gamma\delta^2)-4\pi r^4\alpha\gamma\delta^2-2\pi r^4\gamma^2\delta^2+4\pi\alpha\gamma^2\delta^2r^3\nonumber
\\
f_7&=&6\alpha\gamma^2\delta^2+3r^2(2\alpha+\gamma+\gamma\delta^2)-2r\gamma(\gamma\delta^2+3\alpha+3\alpha\delta^2)\nonumber
\\
\end{eqnarray}

\begin{figure}[h]
\centering
\includegraphics[width=6.5cm]{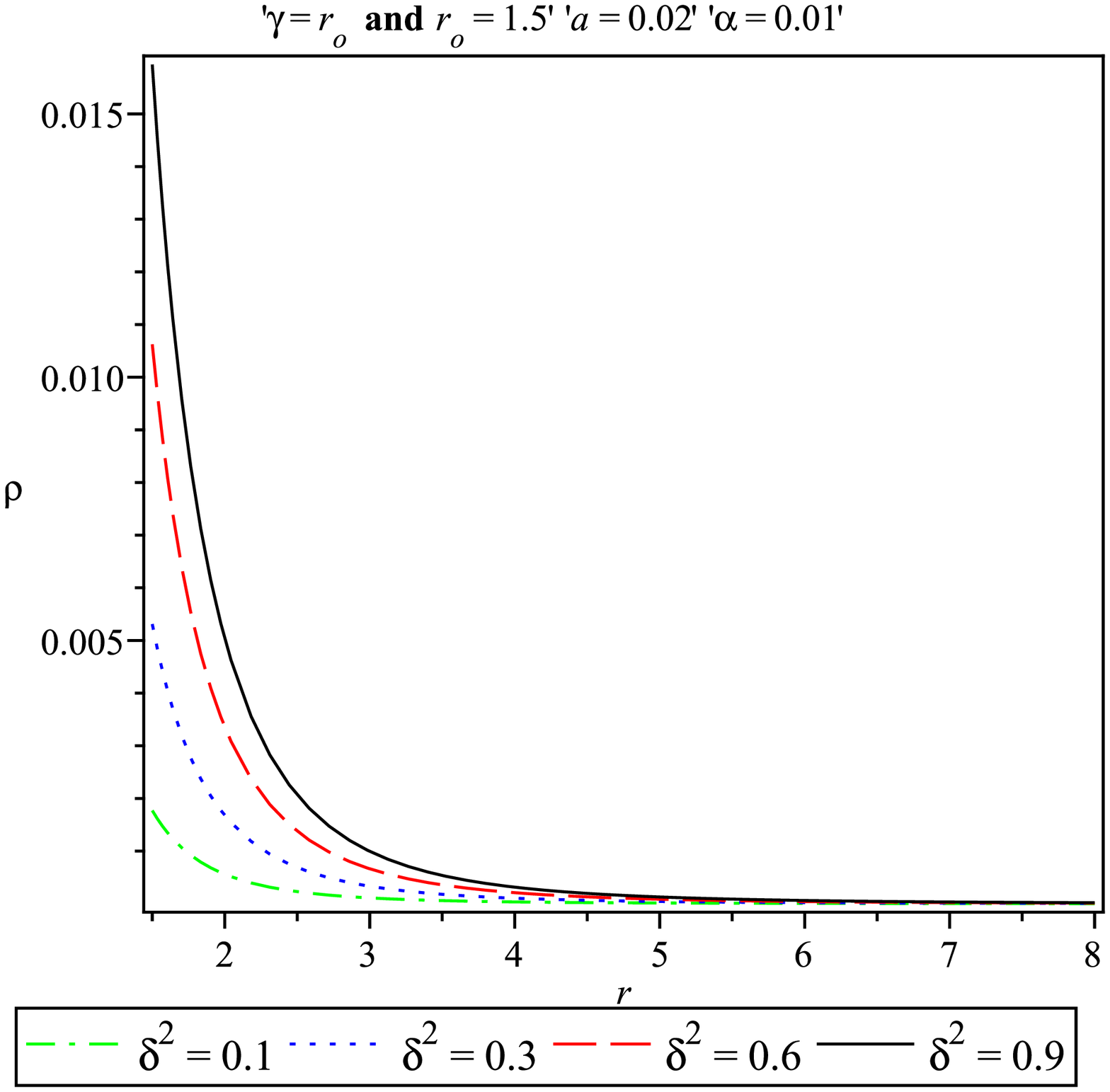}
\includegraphics[width=6.5cm]{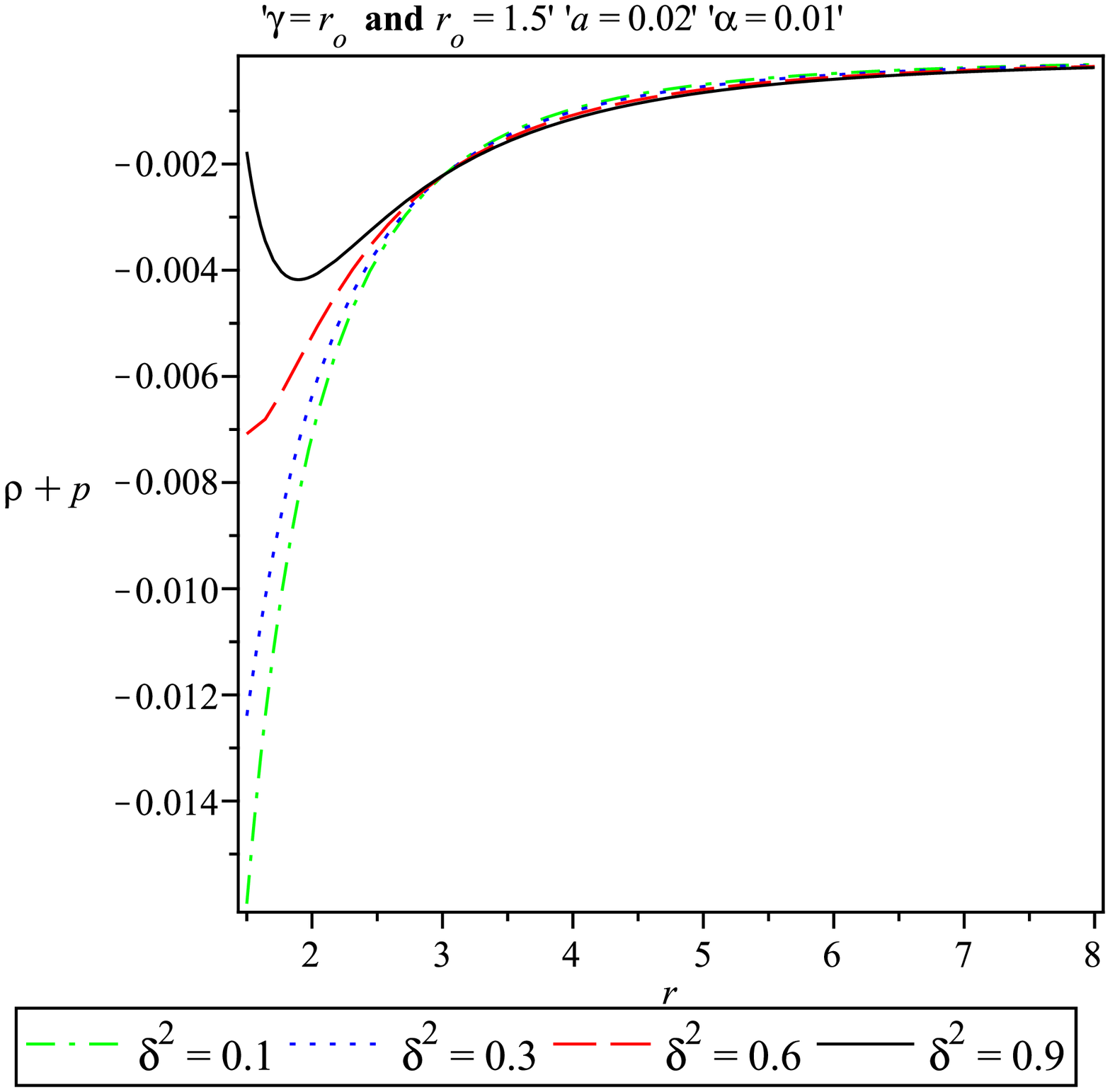} 
\caption{The original matter matter density(Left) and the sum of original matter density and pressure(Right) are   plotted  against the radial coordinate $r ( \geq $ the throat radius) corresponding to $\gamma = r_0 = 1.5$, $a = 0.02$, $\alpha = 0.01$ and $\delta^2 = 0.1, 0.3, 0.6, 0.9$.}\label{fig11}
\end{figure}
\begin{figure}[h]
\centering
\includegraphics[width=6.5cm]{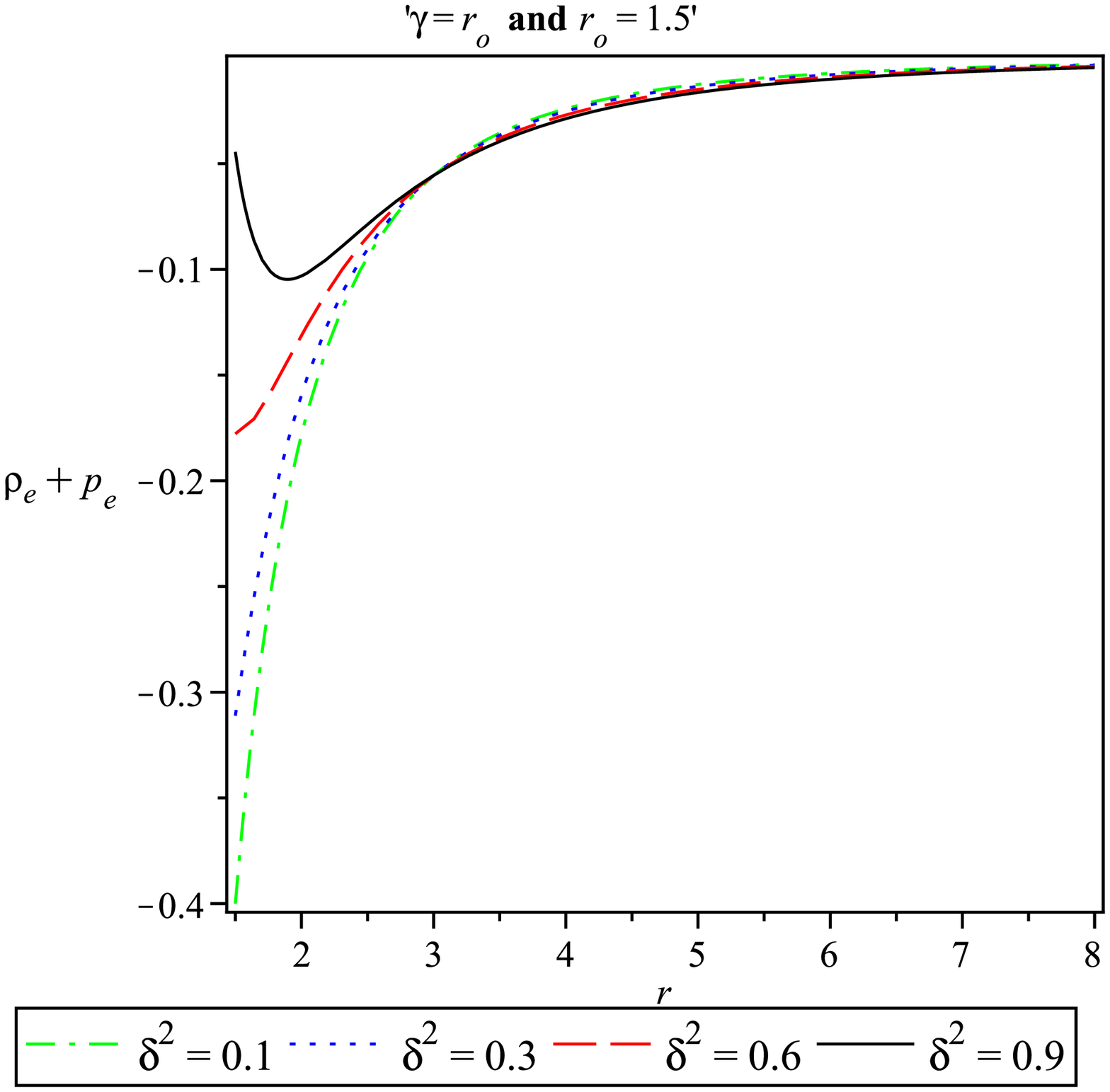}
\includegraphics[width=6.5cm]{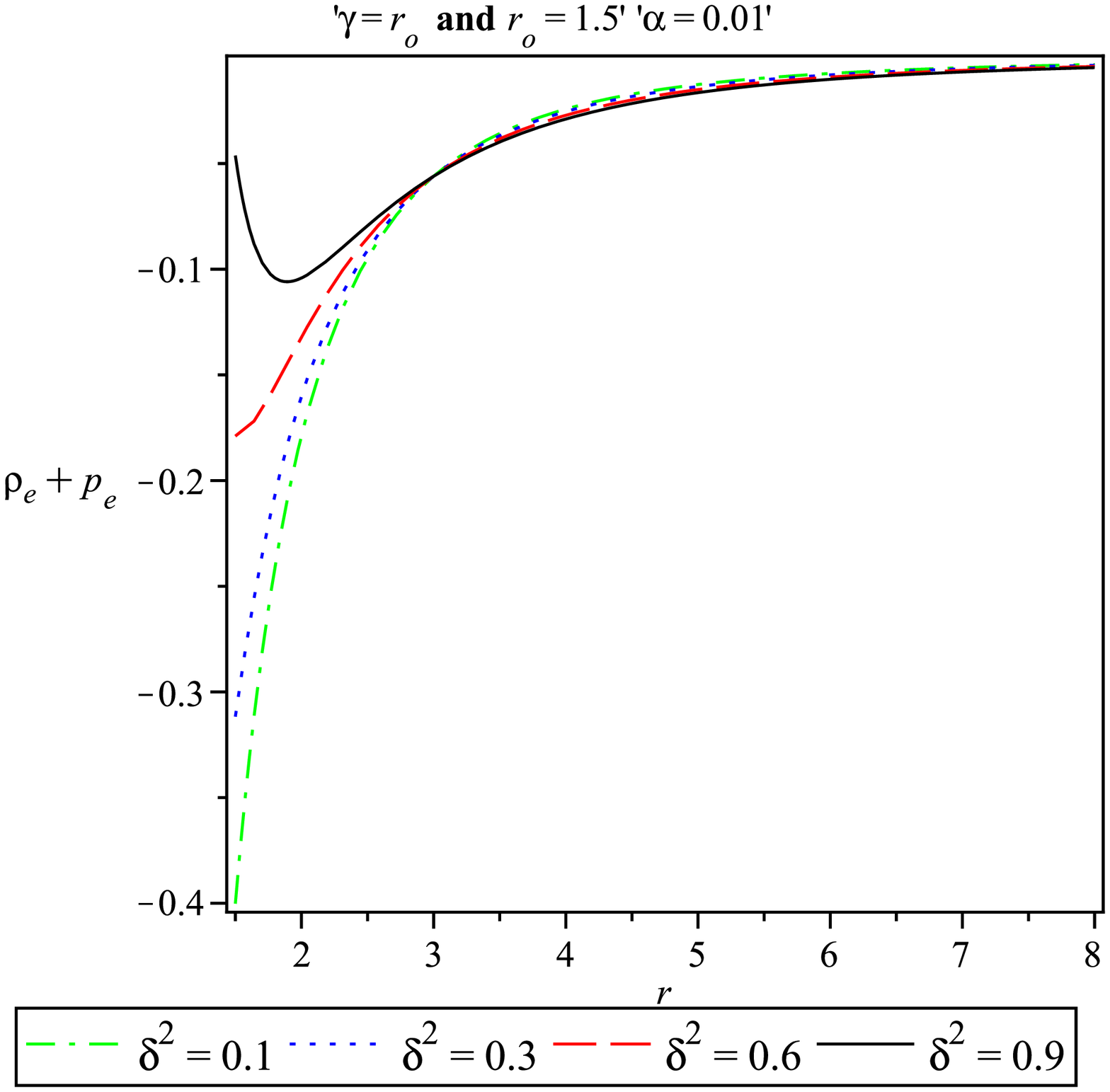}
\caption{The sum of effective matter density and pressure(Left) and the sum of effective matter density and pressure(Right) are   plotted  against  the radial coordinate $r ( \geq $ the throat radius) corresponding to $\gamma = r_0 = 1.5$, $\delta^2 = 0.1, 0.3, 0.6, 0.9$ and $\gamma = r_0 = 1.5$, $\alpha = 0.01$, $\delta^2 = 0.1, 0.3, 0.6, 0.9$, respectively.}\label{fig12}
\end{figure}

In similar argument, anyone can notice that the obtained solutions  for the shape function $ b(r)=\gamma\left\{1+\delta^2\left(1-\frac{\gamma}{r}\right)\right\} $ along with the redshift functions $f(r) = 0$ and $f(r) = \frac{\alpha}{r}$ represent the wormhole structures by violating the null energy condition(NEC), separately (see  Figs.\ref{fig10}(Right),\ref{fig11}(Right) and \ref{fig12}). Moreover, the original matter densities are positive and decreasing  against the radial coordinate $r$, clear from  Figs.\ref{fig10}(Left)-\ref{fig11}(Left). Finally, we can easily see from Eqs.(\ref{rhop22}) and (\ref{rhop14}) that  $\rho(r) + p(r)$ does not tends to zero for any numerical values of $\delta$ and hence there are no possibility to exist the infinitesimal exotic matters that may support the wormholes for these solutions.\\

\section{Results  and Discussions}

In this article, we have found wormholes supporting solutions of the Einstein field equations  in the framework of  $\kappa(R,T)$ gravity. We have obtained the  solutions into two different ways: (i) The solution is obtained by considering a redshift function and a linear equation of state(EoS) and (ii) The solutions are obtained by considering four pairs of two redshift functions and  two shape functions.\\
In the first part, we have taken a redshift function $f(r) = A$, $A$ is constant and a linear equation of state (Eos) $p(r) = m\rho(r)$, where m is a constant along with the restriction $0 < m < 1$ and $ m \neq \frac{1}{3}$. These redshift function and EoS taking into account we have determined the shape function  and the obtained  shape function is positive, decreasing in nature and  satisfied the flare-out condition along with the asymptotic behavior (see Figs.\ref{fig1}-\ref{fig2}(Left)). Therefore,  the shape function satisfied all the necessary conditions to construct a wormhole design and consequently this shape function constructs a wormhole like geometry. The obtained original matter density is positive and decreasing in nature, also the original matter configuration satisfied the weak energy condition(WEC) i.e. $\rho(r) \geq 0$,~$\rho(r) + p(r) \geq 0$  as well as the null energy condition(NEC) i.e. $\rho(r) + p(r) \geq 0$ whereas the modified matter violates the null energy condition(NEC) i.e. $\rho_{e}(r) + p_{e}(r) < 0$), shown in Figs.\ref{fig2}(Right)-\ref{fig3}. Thus, we have obtained a peculiar result that the original matter distribution is the real feasible matter, which provides the fuel to construct and sustain a wormhole in  $\kappa(R,T)$ gravity.\\
Secondly, we have considered four pairs of two different redshift functions and two different shape functions. Both the shape functions are positive, increasing and satisfied all the conditions to present the wormhole structures, shown in Figs.\ref{fig4}, \ref{fig5}(Left), \ref{fig8}(Right) and \ref{fig9}.  The solutions corresponding to the shape function $b(r)=\beta \left(\frac{r}{\beta}\right)^n$ and redshift functions $f(r) = 0,~ f(r) = \frac{\alpha}{r}$ violate the null energy condition, separately, clear from Figs.\ref{fig6}, \ref{fig7}(Right) and \ref{fig8}(Left) i.e. the solutions representing matter distributions hold the wormhole structures in $\kappa(R,T)$ gravity. Moreover, an arbitrarily small amount of exotic matter is responsible to construct a wormhole design for these solutions. Also, the matter distributions represented by the solutions corresponding to the shape function $b(r)=\gamma\left\{1+\delta^2\left(1-\frac{\gamma}{r}\right)\right\}$ and redshift functions $f(r) = 0,~ f(r) = \frac{\alpha}{r}$ retain the wormhole structures by violating the null energy condition(NEC) (see Figs.\ref{fig10}(Right), \ref{fig11}(Right) and \ref{fig12}), separately but there are no possibility to exist the infinitesimal exotic matters that may support the wormholes.

{Now it will be very interesting to present a comparative analysis with the other WH works in modified theories of gravitation. Godani and Samanta \cite{go19} have investigated the traversable WHs and energy conditions within the framework of $f(R)$ modified theory gravity. They have used the shape function of the form $b(r)=\beta\left(\frac{r}{\beta}\right)^{n}$, and they have found that the energy density is positive and all energy conditions are violated which supports the existence of WHs. On the other hand, Samanta et al. \cite{sam18} have presented the WHs with exponential shape function within the framework of $f(R)$ gravity. They have concluded that the WH may not contain same type of matter for all values of $r$ and it is filled some abnormal type of matter, which is may be called as dark energy. But, in our case, the WHs in $\kappa(R,T)$ gravity violates the energy conditions for all values of $r$ and WH contains same type of matter. Hence, our results are coincide with the results obtained by Godani and Samanta \cite{go19}.}

\subsection*{Acknowledgments}
Farook Rahaman  would like to thank the authorities of the Inter-University Centre for Astronomy
and Astrophysics, Pune, India for providing the research facilities. Susmita Sarkar and  Nayan Sarkar are
thankful to UGC (Grant No.: 1162/(sc)(CSIR-UGC NET , DEC 2016)) and CSIR (Grand No.-09/096(0863)/2016-EMR-I.), Govt. of India  for financial support respectively.

\end{document}